\documentclass[12pt]{article}
\usepackage{latexsym}
\usepackage{amssymb}
\catcode`\@=11
\newcount\hour
\newcount\minute
\newtoks\amorpm \hour=\time\divide\hour by 60\minute
=\time{\multiply\hour by 60 \global\advance\minute by-\hour}
\edef\standardtime{{\ifnum\hour<12 \global\amorpm={am}%
        \else\global\amorpm={pm}\advance\hour by-12 \fi
        \ifnum\hour=0 \hour=12 \fi
        \number\hour:\ifnum\minute<10
        0\fi\number\minute\the\amorpm}}
\edef\militarytime{\number\hour:\ifnum\minute<10
0\fi\number\minute}
\def\draftlabel#1{{\@bsphack\if@filesw {\let\thepage\relax
   \xdef\@gtempa{\write\@auxout{\string
      \newlabel{#1}{{\@currentlabel}{\thepage}}}}}\@gtempa
   \if@nobreak \ifvmode\nobreak\fi\fi\fi\@esphack}
        \gdef\@eqnlabel{#1}}
\def\@eqnlabel{}
\def\@vacuum{}
\def\marginnote#1{}
\def\draftmarginnote#1{\marginpar{\raggedright\scriptsize\tt#1}}
\def\draft{
        \pagestyle{plain}
        \overfullrule=2pt
        \oddsidemargin -.1truein
        \def\@oddhead{\sl \phantom{\today\quad\militarytime} \hfil
        \smash{\Large\sl DRAFT} \hfil \today\quad\militarytime}
        \let\@evenhead\@oddhead
        \let\label=\draftlabel
        \let\marginnote=\draftmarginnote
        \def\ps@empty{\let\@mkboth\@gobbletwo
        \def\@oddfoot{\hfil \smash{\Large\sl DRAFT} \hfil}
        \let\@evenfoot\@oddhead}
        \def\@eqnnum{(\theequation)\rlap{\kern\marginparsep\tt\@eqnlabel}%
        \global\let\@eqnlabel\@vacuum}  }

\renewcommand{\theequation}{\thesection.\arabic{equation}}
\renewcommand{\thefootnote}{\fnsymbol{footnote}}
\newcommand{\newsection}{    
\setcounter{equation}{0}\section}
\def\appendix#1{\addtocounter{section}{1}\setcounter{equation}{0}
\renewcommand{\thesection}{\Alph{section}}
\section*{Appendix \thesection\protect\indent \parbox[t]{11.15cm}{#1}}
\addcontentsline{toc}{section}{Appendix \thesection\ \ \ #1}}

\def \lc {{light-cone}}

\def \bi{\bibitem}
\def \la {\label}

\def \b {\beta}
\def \Om {\Omega}

\def \d {\partial}

\def\be{\begin{equation}}
\def\ee{\end{equation}}

\catcode`\@=11

\def\bea{\begin{eqnarray}}
\def\eea{\end{eqnarray}}
\def\beann{\begin{eqnarray*}}
\def\eeann{\end{eqnarray*}}
\def\beq{\begin{equation}}
\def\eeq{\end{equation}}
\def\ba{\begin{array}}
\def\ea{\end{array}}
\def\ben{\begin{enumerate}}
\def\een{\end{enumerate}}

 \def \la {\label}
 \def\be{\begin{equation}}
\def\ee{\end{equation}}

\def \la {\label}

\font\mybb=msbm10 at 11pt

\def\bb#1{\hbox{\mybb#1}}

\def\bR {\bb{R}}

\def\bC {\bb{C}}

\def\e  {\epsilon}

 \def\ep {\epsilon}

\def \ee {\epsilon}

\def \g {\gamma}
\def \bi{\bibitem}
\def\a{\alpha }

\def \ep {\epsilon}

\def \d {\delta}

\def \g {\gamma}

\def \b {\beta}

\def\lc{\lrcorner}

\def\be{\begin{equation}}
\def\ee{\end{equation}}

\def \bi {\bibitem}
\def \la{\label}

\begin{document}
\date{November 2002}
\begin{titlepage}
\begin{center}

\vspace{2.0cm}
{\Large \bf  The spinorial geometry of supersymmetric IIB backgrounds}\\[.2cm]

\vspace{1.5cm}
 {\large  U. Gran$^1$, J. Gutowski$^2$ and  G. Papadopoulos$^1$}

 \vspace{0.5cm}
${}^1$ Department of Mathematics\\
King's College London\\
Strand\\
London WC2R 2LS, UK\\
\vspace{0.5cm}
${}^2$ Mathematical Institute\\
Oxford University\\
Oxford OX1 3LB, UK
\end{center}

\vskip 1.5 cm
\begin{abstract}

We investigate the Killing spinor equations of IIB supergravity for one
Killing spinor. We show that there are three types of orbits
of $Spin(9,1)$ in the space of Weyl spinors which give rise to Killing  spinors with
stability subgroups $Spin(7)\ltimes \bR^8$, $SU(4)\ltimes \bR^8$
and  $G_2$. We solve the Killing spinor equations for the $Spin(7)\ltimes \bR^8$ and
 $SU(4)\ltimes \bR^8$ invariant spinors, give the fluxes in terms of the geometry
and determine the conditions on the spacetime geometry imposed by
supersymmetry. In both cases, the spacetime  admits a null, self-parallel,  Killing
 vector field.
We also apply our formalism to  examine a class of
 $SU(4)\ltimes \bR^8$ backgrounds
 which admit one and two pure spinors as Killing spinors and investigate
 the geometry of the spacetimes.

\end{abstract}
\end{titlepage}
\newpage
\setcounter{page}{1}
\renewcommand{\thefootnote}{\arabic{footnote}}
\setcounter{footnote}{0}

\setcounter{section}{0}
\setcounter{subsection}{0}
\newsection{Introduction}


In the last few years there has been much interest in the systematic understanding
of supersymmetric solutions of ten- and eleven-dimensional supergravities.
The maximal supersymmetric solutions of eleven-dimensional
supergravity have been classified in \cite{jfgpa, jfgpb} by exploring the
vanishing of the curvature of the supercovariant connection of the theory.
The Killing spinor equations have also been solved for one
Killing spinor and the geometry of the spacetime has been investigated
in \cite{pakis, gutowski}. This
has been done by using the properties of spinor bilinears.
The supersymmetric backgrounds of eleven-dimensional supergravity
have also be examined  using the holonomy of the supercovariant connection
\cite{duff, hull, tsimpisa}, see also \cite{bandos} and \cite{liu}. In \cite{josec}
it has been shown that
that backgrounds with more than twenty four supersymmetries are locally homogeneous
spaces.

Recently, a new method for solving the Killing spinor equations
of supergravity theories has been proposed in \cite{gran} and applied
to eleven-dimensional supergravity. This is
based on a realization of spinors in terms of forms and the introduction
of a  basis in the space of spinors.
Using this method, one can easily    analyze the Killing spinor equations
and  determine the geometry of the associated spacetime.
As a demonstration of the effectiveness of this method,
the Killing spinor equations of eleven-dimensional
supergravity have been solved for one, two, three
and four spinors with stability subgroups
$SU(5)$ and $SU(4)$ \cite{gran}.

Some progress towards a systematic understanding of
supersymmetric solutions of IIB supergravity \cite{schwarz, howe} has also been made.
The maximal supersymmetric solutions of IIB supergravity
been classified in \cite{jfgpa, jfgpb}. It has been found that they are
locally isometric to Minkowski space, $AdS_5\times S^5$ \cite{schwarz} and the maximal
supersymmetric plane wave \cite{georgea}, and they are related by  Penrose limits
 \cite{georgeb}.
 The holonomy of the the supercovariant connection of IIB supergravity
is $SL(32,\bR)$  which reduces for backgrounds with $N$ spinors to a subgroup
of $SL(32-N, \bR)\ltimes \oplus_N\bR^{32-N}$ \cite{tsimpis}.
 In addition,
the Killing spinor equations of IIB supergravity have been expressed
as the parallel transport equations for the associated form bi-linears
in \cite{jones}.  Other methods have also been used to solve the IIB Killing spinor
equations like for example the `algebraic spinor' technique  which
has been applied to construct supersymmetric flows \cite{warner}.

In this paper, we use the method proposed in \cite{gran} to solve
the Killing spinor equations of IIB supergravity for backgrounds
that admit  a  $Spin(7)\ltimes\bR^8$  or  an
$SU(4)\ltimes\bR^8$ invariant  Killing spinor. As an application, we  examine
backgrounds that admit  one and  two $SU(4)\ltimes\bR^8$ invariant pure Killing
spinors. To apply the method, one  has to take the following steps.
\begin{itemize}
\item Find a realization of spinors in terms of forms and construct
a basis in the space of spinors.

\item Find a canonical or normal form for the Killing spinors up to the
gauge transformations of the Killing spinor equations of the supergravity theory.

\item Substitute the canonical form of
 Killing spinors into the Killing spinor equations and use the
basis in the space of spinors to turn the Killing spinor equations
 into a linear system for the fluxes,
the geometry and the spacetime derivatives of  functions that
determine locally the Killing spinors.

\item Solve the linear system for the fluxes
 and find the conditions that arise
on the geometry of the spacetime\footnote{The functions that
 the Killing spinors depend on may also be restricted.}.

\end{itemize}

A description of $Spin(9,1)$  spinors in terms of forms is presented in appendix A.
A suitable  basis in the space of spinors for our analysis is also  given.
The manifest gauge invariance of the IIB
supergravity depends on the formulation of the theory. If the supercovariant
connection is written as in \cite{schwarz}, then the gauge invariance
is $Spin(9,1)\times U(1)$, where $U(1)$ is a local (duality) gauge
group\footnote{The field equations of IIB supergravity
have an additional global $SL(2,\bR)$
symmetry \cite{west, schwarz}}. However, it is convenient not to fix the $U(1)$ part
of the gauge symmetry because it may be  used  later
to simplify computations for specific backgrounds, see \cite{granb}.
Because of this, we use only the  $Spin(9,1)$
gauge group to bring the Killing  spinor into a canonical form.
We find  that there are
three cases to be considered which are distinguished by the stability
subgroup of the Killing spinors in $Spin(9,1)$. These are $Spin(7)\ltimes \bR^8$,
$SU(4)\ltimes \bR^8$ and $G_2$. The canonical
forms of the Killing spinors written in terms of forms are
\bea
\epsilon&=& (f+ig) (1+  e_{1234})~,
\cr
\epsilon &=& (f-g_2+ig_1) 1+ (f+g_2+ig_1)e_{1234}~,
\cr
\epsilon &=& f(1+e_{1234})+ i {g\over\sqrt{2}} \Gamma^+ (e_1+e_{234})~,
\la{caform}
\eea
respectively, where $f, g, g_1, g_2$ are real functions of the spacetime.
The Killing spinors depend on more than one spacetime function.

We then substitute the $Spin(7)\ltimes \bR^8$ and
$SU(4)\ltimes \bR^8$ invariant spinors into the Killing spinor equations\footnote{The
$G_2$ invariant case will be investigated elsewhere \cite{gju}.}.
In both cases, the linear system that
we derive, after expanding the Killing spinors in the spinor basis that we have
constructed in appendix A, is rather involved. However {\it all} equations can be solved
 to express some of the fluxes in terms of the
geometry and to find the conditions on the geometry imposed by supersymmetry.
The expressions for the fluxes are simplified using the self-duality condition
of the five-form flux $F$.

The conditions on the  geometry, which are expressed as relations
between components of the Levi-Civita
connection of the spacetime, can be  directly analyzed to specify the geometry of the
spacetime.
For the $Spin(7)\ltimes \bR^8$ and $SU(4)\ltimes \bR^8$ cases, it is also convenient
to consider the spacetime form bi-linears  associated with the Killing spinor.
 However unlike the eleven-dimensional
supergravity case, the spinors of IIB supergravity are complex.
Because of this acting on the Killing spinors with Pin and Spin
invariant operations, one can construct new spinors that are
defined on the spacetime. One such operator is $C *$, where $*$ is
the standard complex conjugation and $C$ is a charge conjugation matrix (see appendix A).
Therefore if $\epsilon$ is a Killing spinor, then $\tilde \epsilon=C(\epsilon^*)$
is a spinor defined
on the spacetime but not always Killing. It
turns out that for the geometric interpretation of the conditions
on the Levi-Civita connection arising from the Killing spinor
equations, it is necessary to construct  the spacetime form
bi-linears of the pairs $(\epsilon, \epsilon)$, $(\epsilon, \tilde\epsilon)$
and $(\tilde\epsilon, \tilde\epsilon)$.
The spacetime of supersymmetric IIB backgrounds with
$Spin(7)\ltimes \bR^8$ and $SU(4)\ltimes \bR^8$ invariant Killing spinors
admits a {\it null}, {\it self-parallel}, {\it Killing} vector field.
In addition the  associated
spacetime admits a $Spin(7)\ltimes \bR^8$ and $SU(4)\ltimes \bR^8$
structure, respectively.  Examples
of such spacetimes are Lorentzian extensions of one-parameter families
of eight-dimensional manifolds with generic $Spin(7)$ and $SU(4)$
structures.

A special  class of IIB backgrounds with $SU(4)\ltimes
\bR^8$-invariant parallel spinors are those for which the Killing
spinor is  pure. Pure spinors have been considered before in
relation to the Killing spinor equations of IIB supergravity in
\cite{minasian} in a somewhat different context. The definition of
pure spinor can be given in different ways. One way\footnote{Another
way is to say that the pure spinor is annihilated by a maximal
isotropic subspace of the spacetime, i.e. half of the gamma matrices
along some directions annihilate the spinor.} is to assert that a
spinor is pure iff the one-form bi-linear of the spinor vanishes.
Using this definition, one can find that an $SU(4)\ltimes
\bR^8$-invariant spinor is pure if either
\bea
\epsilon &=& h\, 1~,
\eea
i.e. $g_1=0$ and $f=-g_2=h/2$,
or
\bea
\epsilon &=& k\, e_{1234}~,
\eea
i.e. $g_1=0$ and $f=g_2=k/2$~.
We shall analyze the conditions in both cases. In particular, we shall
investigate the $\epsilon =h 1$ case in some detail. This is  because it is rather
straightforward to solve the Killing spinor equations and it has most of the
features of the generic $SU(4)\ltimes \bR^8$ invariant case.
We shall also summarize the conditions required by supersymmetry for
the $\epsilon = k\, e_{1234}$  Killing spinor.
Then, we shall give the conditions required for  both $h 1$ and $k e_{1234}$ to be
Killing spinors and we shall investigate the geometry of the associated
spacetime.

The $Spin(7)\ltimes \bR^8$ invariant Killing spinor  is a special case of
the $SU(4)\ltimes \bR^8$ invariant one, as it can be seen by setting
 $g_2=0$ and $g_1=g$ in (\ref{caform}).
Nevertheless it turns out that some of the conditions that arise on the geometry
and some of the expressions
for the fluxes are different. Because of this, we shall treat
them as two distinct cases. We shall point out some of the differences in the
geometry at the relevant sections.

This paper has been organized as follows: In section two, we use the
$Spin(9,1)$ gauge transformations of the Killing spinor equations of
IIB supergravity to bring the Killing spinors to a canonical form.
We also present the IIB Killing spinor equations and investigate
some of their properties. In section three, we give the linear
system that arises from the Killing spinor equations and   summarize
the conditions on the geometry of the spacetime that admits an
$SU(4)\ltimes \bR^8$ invariant Killing spinor. We also investigate
these conditions using the spacetime-form bi-linears that are
associated to the Killing spinor. In appendix B, we explain the
derivation of the linear system and express the fluxes in terms of
the geometry. In section four, we summarize the conditions on the
geometry of the spacetime that admits a $Spin(7)\ltimes \bR^8$
invariant Killing spinor and investigate the geometry of the
associated spacetime. In appendix C, we express the fluxes in terms
of the geometry for the $Spin(7)\ltimes \bR^8$ invariant Killing
spinor. In section five, we present the conditions on the geometry
for backgrounds with  $h 1$ and $k e_{1234}$ pure Killing spinors.
The fluxes for these cases are given in appendices D and E. In
section six, we solve the Killing spinor equations for backgrounds
which admit  both $h1$ and $ k e_{1234}$ as Killing spinors and
investigate the geometry of the associated spacetimes.

\newsection{Orbits of $Spin(9,1)$ and Killing spinor equations}

\subsection{Orbits of $Spin(9,1)$ in $\Delta_{16}^+$}

 The group $Spin(9,1)$ has one type of orbit
with stability subgroup $Spin(7)\ltimes \bR^8$ in the Majorana-Weyl
representation\footnote{Our spinor conventions as well as the realization
of spinors in terms of forms is explained in detail in appendix A.} $\Delta_{16}^+$.
To see this, consider the spinor represented by
\be
1+e_{1234}~.
\ee
The stability subgroup of this spinor in $Spin(9,1)$ is $Spin(7)\ltimes \bR^8$.
This can be easily seen by adapting a computation
of \cite{jose} done for eleven-dimensional supergravity to this case,
see also \cite{bryant} and \cite{sfetsos}. To find the stability subgroup, we
solve the infinitesimal equation
\be
\lambda_{AB} \Gamma^{AB} (1+ e_{1234})=0~,
\ee
where $\lambda$ parameterizes  the infinitesimal  spinor transformations.
This computation is most easily done in the pseudo-Hermitian basis
 (\ref{hbasis}).
It is easy to see that the above condition implies
that the parameters are restricted as
\bea
\lambda_{\bar\a\bar\b}={1\over2} \epsilon_{\bar\a\bar\b}{}^{\g\d}  \lambda_{\g\d}~,~~~~~
\lambda_{\a\bar\b} g^{\a\bar\b}=\lambda_{-+}=\lambda_{+\a}=\lambda_{+\bar\a}=0~,
\la{parasp}
\eea
where $\epsilon_{\bar1\bar2\bar3\bar4}=1$.
Observe that  $\lambda_{-\a}$ and $\lambda_{-\bar\a}$
are complex conjugate to each other
but otherwise unconstrained. It is known that the Lie algebra
$spin(7)$ in a Hermitian basis is spanned by traceless (1,1)-forms and
(2,0)-forms in $\bC^4$ which are related to their complex conjugates by a duality relation
as in the first equation of (\ref{parasp}), see e.g. \cite{salamon}.  Therefore
the group that
leaves invariant $1+e_{1234}$ has Lie algebra
$spin(7)\oplus \bR^8$.  So we shall take
 the stability subgroup\footnote{There may be subtleties with discrete groups,
see \cite{amus, mcinnes}.}  to be $Spin(7)\ltimes \bR^8$. Note that the product
is semi-direct because $Spin(7)$ acts on $\bR^8$ with a spin representation.

Having established this, we decompose $\Delta_{16}^+$ under the stability
subgroup $Spin(7)$ as
\be
\Delta_{16}^+= \bR<1+e_{1234}>+\Lambda^1(\bR^7)+\Delta_8~,
\la{dec}
\ee
where the singlet $\bR$ is generated by $1+e_{1234}$, $\Lambda^1(\bR^7)$
is the vector representation of $Spin(7)$ which is spanned
by the Majorana spinors associated with two-forms in the directions $e_1, \dots, e_4$ and
$i(1-e_{1234})$, and $\Delta_8$ is the spin representation of $Spin(7)$
which is spanned by the rest of Majorana spinors which are of the type
 $\Gamma^+\eta$, $\eta$ is a  spinor  generated by the odd forms in
 the directions $e_1, \dots, e_4$.
Therefore the most general spinor in $\Delta_{16}^+$ can be written as
\be
\eta= a(1+e_{1234})+ \theta_1+\theta_2~,
\ee
where $\theta_1\in \Lambda^1(\bR^7)$ and $\theta_2\in \Delta_8$.
First we assume that $a\not=0$. In this case, there are
 two cases to consider depending on whether $\theta_2$ vanishes or not.
If $\theta_2=0$, since $Spin(7)$ acts with the vector representation
on $\Lambda^1(\bR^7)$, it is always possible to choose $\theta_1=i b(1-e_{1234})$.
The most general spinor in this case then is
\be
\eta= a (1+e_{1234})+ i b (1-e_{1234})~.
\ee
However, it is easy to see that this spinor is in the same orbit as $1+e_{1234}$, e.g.
observe that
\be
\eta= h e^{\psi \Gamma_{16}} (1+e_{1234})~,
\ee
where $h^2= a^2+b^2$ and $\tan\psi= b/a$. Next suppose that $\theta_2$ does not vanish.
If $\theta_2\not=0$, there is always a $Spin(7)$ transformation such that
$\theta_2= c \Gamma^+ (e_1+e_{234})$. This is because
$Spin(7)$ acts transitively on the $S^7$ in $\Delta_8$ and the stability subgroup
is $G_2$, $Spin(7)/G_2=S^7$.
In addition $G_2$ acts transitively on the $S^6$ in
$\Lambda^1(\bR^7)$ with stability subgroup $SU(3)$, see e.g. \cite{salamon}.
So it can always be arranged such that $\theta_1=i b (1-e_{1234})$.
Therefore the most general spinor in this case is
\be
\eta=a (1+e_{1234})+ i b (1-e_{1234})+ c \Gamma^+ (e_1+ e_{234})~.
\ee
However observe that this spinor is in the same orbit as $1+e_{1234}$. Indeed
\be
\eta= e^{{b\over 2c} \Gamma^- \Gamma^6} e^{{c\over a} \Gamma^+ \Gamma^1}
a (1+e_{1234})~.
\ee
So, we conclude that if $a\not=0$, then there is one orbit represented by
$a (1+e_{1234})$. It remains to investigate the case where
$a=0$. In this case, it is straightforward to see that the orbit
can always be represented by $c \Gamma^+ (e_1+e_{234})$. In turn,
this spinor is in the same orbit of $Spin(9,1)$ as ${c\over \sqrt{2}} (1+e_{1234})$
as  can seen
by acting on the latter with the element $\Gamma_5\Gamma_1$ of $Spin(9,1)$.
As a consequence, the stability subgroup of $c \Gamma^+ (e_1+e_{234})$
is again  $Spin(7)\ltimes \bR^8$. Therefore we conclude that there is
only one type of orbit of $Spin(9,1)$ in $\Delta^+_{16}$ which can be
represented with $a(1+e_{1234})$.

In IIB supergravity, the Killing spinors are (complex) Weyl and
so they take values in two copies of the same Majorana-Weyl representation $\Delta_{16}^+$.
To find the most general Killing spinor
that can arise in the theory, we assume that the Killing spinor
in the first copy is represented by $a(1+e_{1234})$ and
 decompose the second Majorana-Weyl representation under the stability
subgroup $Spin(7)\ltimes \bR^8$ of $\eta_1$.
As we have mentioned, $\Delta_{16}^+$ under  $Spin(7)$ decomposes as (\ref{dec}).
Using the same arguments as those below (\ref{dec}), we can choose the two
Majorana Weyl spinors of IIB supergravity to take the  form
\bea
\eta_1&=& a(1+e_{1234})~,
\cr
\eta_2&=& b_1 (1+e_{1234})+ i b_2 ( 1-e_{1234})+ b_3 \Gamma^+ (e_1+e_{234})~.
\la{iibasp}
\eea
So far we have been concerned with the orbits of $Spin(7)$. At this point
we have to distinguish
between two different cases. First suppose that $b_3\not=0$. In this case,
we can further
consider the action of  $\bR^8$ on the (\ref{iibasp}) spinors.
In particular observe that
\be
\eta_2=e^{-{b_1\over2 b_3} \Gamma^-\Gamma^6+{b_2\over2 b_3} \Gamma^-\Gamma^1}
b_3 \Gamma^+ (e_1+e_{234})~.
\ee
Therefore, we have shown that in this case the two Majorana-Weyl
spinors of IIB supergravity can be represented by
$\eta_1= a(1+e_{1234})$ and $\eta_2= b \Gamma^+ (e_1+e_{234})$.
The  canonical form of the spinors is
\bea
\eta_1&=& f(1+e_{1234})~,
\cr
\eta_2&=& g \Gamma^+ (e_1+e_{234})~,
\la{iibsp}
\eea
where $f, g$ are real spacetime functions. In particular, the Killing spinor
in (\ref{kseqna}) and ({\ref{kseqnb}}) below is $\epsilon=\eta_1+i \eta_2$.
The stability subgroup of the  Killing spinor  is $G_2$.

Next suppose that $b_3=0$. In this case, we have
\bea
\eta_1&=& f(1+e_{1234})~,
\cr
\eta_2&=& g_1 (1+e_{1234})+ i g_2 ( 1-e_{1234})~,
\la{iibaspa}
\eea
where $f, g_1$ and $g_2$ are real spacetime functions. The Killing spinor
in
(\ref{kseqna}) and ({\ref{kseqnb}}) below is
 $\epsilon=\eta_1+i\eta_2$.
The stability subgroup of the Killing spinor is $SU(4)\ltimes \bR^8$.

One can also take $b_3=b_2=0$, in this case we have
\bea
\eta_1&=& f(1+e_{1234})~,
\cr
\eta_2&=& g (1+e_{1234})~,
\la{iibbsp}
\eea
where $f,g$ are real spacetime functions. Similarly, the
Killing spinor
in
(\ref{kseqna}) and ({\ref{kseqnb}}) below is
 $\epsilon=\eta_1+i\eta_2$ and has stability subgroup $Spin(7)\ltimes \bR^8$.

To summarize, there are three types of orbits of $Spin(9,1)$ in the Weyl spinor
representation with stability subgroups $Spin(7)\ltimes \bR^8$, $SU(4)\ltimes \bR^8$
and $G_2$. The representatives of these orbits are given in  (\ref{caform}).
Unlike the M-theory case, the Killing spinor depend on more than one spacetime function.
It is straightforward to extend the analysis in this section  to
 backgrounds which
have more than one supersymmetry.

\subsection{ Killing spinor equations}

The bosonic fields of IIB supergravity are the spacetime metric $g$, two real scalars,
 the axion $\sigma$
and the dilaton $\phi$, two three-form field strengths $G_1$ and $G_2$, and a self-dual
five-form field strength $F$. The  Killing spinor
equations of IIB supergravity are the parallel transport equations of the
supercovariant derivative ${\cal D}$
\cite{schwarz}
\be
\label{eqn:gravitino}
{\cal D}_M\epsilon=\tilde\nabla_M \epsilon+{i\over 480} \Gamma^{N_1\dots N_5} \Gamma_M\epsilon
 F_{N_1\dots N_5}
 -{1\over 96} (\Gamma_{M}{}^{N_1N_2N_3}
G_{N_1N_2N_3}-9 \Gamma^{N_1N_2} G_{MN_1N_2})(C\epsilon)^*=0~,
\ee
and the algebraic condition
\be
\label{eqn:alg}
P_M \Gamma^M (C\epsilon)^*+ {1\over 24} G_{N_1N_2N_3} \Gamma^{N_1N_2N_3} \epsilon=0~,
\ee
where
$$
\tilde\nabla_M=D_M+{1\over4} \Omega_{M,AB} \Gamma^{AB}~,~~~~~~D_M=\partial_M-{i\over2}Q_M
$$
is the
spin connection, $\nabla_M=\partial_M+{1\over4} \Omega_{M,AB} \Gamma^{AB}$,
twisted with $U(1)$ connection $Q_M$, $Q_M^*=Q_M$, $\epsilon$
is a (complex) Weyl spinor,
$\Gamma^{0\dots 9}\epsilon=-\epsilon$, and  $C$ is a charge conjugation
 matrix\footnote{In the basis of gamma matrices
chosen in \cite{schwarz}, $C=1$, and so it has been neglected,
see however \cite{becker}.}. (For our spinor conventions see appendix A.)
  Killing spinor equations are the  vanishing conditions of the supersymmetry
transformations of the gravitino, and  the supersymmetric partners of the
dilaton and axion  restricted to the bosonic sector of IIB supergravity, respectively.
The precise dependence of the complex field strengths $G$ and $P$ on the scalars
and $G_1,G_2$
field strengths is described\footnote{We use a mostly plus
convention for the metric.
To relate this  to the conventions of  \cite{schwarz}, one
takes  $\Gamma^A\rightarrow
i\Gamma^A$ and every time a index is lowered there is also
an additional minus sign.}
 in \cite{schwarz} and we shall not repeat the formulae here.
 For a superspace formulation of IIB supergravity
 see \cite{howe}.

It is convenient to choose the orientation of spacetime as $\epsilon_{01\dots9}=1$.
In this case, the self-duality
condition on $F$ is  $F_{M_1\dots M_5}=- {1\over 5!}
\epsilon_{M_1\dots M_5}{}^{N_1\dots N_5}
F_{N_1\dots N_5}$. Because $\epsilon$ is chiral,
the Killing spinor equations can be rewritten as
\be
\tilde\nabla_M \epsilon+{i\over 48} \Gamma^{N_1\dots N_4 } \epsilon
 F_{N_1\dots N_4 M}
 -{1\over 96} (\Gamma_{M}{}^{N_1N_2N_3}
G_{N_1N_2N_3}-9 \Gamma^{N_1N_2} G_{MN_1N_2}) (C\epsilon)^*=0
\la{kseqna}
\ee
and
\be
P_M \Gamma^M (C\epsilon)^*+ {1\over 24} G_{N_1N_2N_3} \Gamma^{N_1N_2N_3} \epsilon=0~.
\la{kseqnb}
\ee
Observe that the above Killing spinor equations are at most fourth order in the gamma
matrices.

It is worth exploring some general properties of the above Killing spinor equations.
Acting with the anti-linear map $C*$,  the Killing spinor equations
(\ref{kseqna}) and (\ref{kseqnb}) become
\bea
\nabla_M (C\epsilon)^*+{i\over2} Q_M (C\epsilon)^*
 -{i\over 48} \Gamma^{N_1\dots N_4 } (C\epsilon)^*
 F_{N_1\dots N_4 M}
 \cr
 -{1\over 96} (\Gamma_{M}{}^{N_1N_2N_3}
G^*_{N_1N_2N_3}-9 \Gamma^{N_1N_2} G^*_{MN_1N_2}) \epsilon=0~,
\la{kseqnac}
\eea
where $\nabla$ is the spin connection and $Q_M^*=Q_M$,
and
\be
P^*_M \Gamma^M \epsilon+ {1\over 24} G^*_{N_1N_2N_3} \Gamma^{N_1N_2N_3} (C\epsilon)^*=0~,
\la{kseqnbc}
\ee
respectively.
Comparing (\ref{kseqna}), (\ref{kseqnb}) and (\ref{kseqnac}),
(\ref{kseqnbc}), it is clear that  if $\epsilon$
is a parallel spinor, then $(C\epsilon)^*$ is also a parallel
spinor provided that the fluxes $P, G$ are real
and $F=0$. In such a case, the background will have at least two parallel spinors.

Next suppose that  $\epsilon$ is a Majorana-Weyl spinor.
The Majorana condition implies that   $(C\epsilon)^*=\epsilon$.
This can be used  to rewrite the Killing spinor equations
(\ref{kseqna}) and (\ref{kseqnb}), and (\ref{kseqnac}) and
(\ref{kseqnbc}),  in terms of $\epsilon$ only.
 Taking the difference of (\ref{kseqna}) and (\ref{kseqnac}), we find
 \bea
 -i Q_M\epsilon+{i\over 24} \Gamma^{N_1\dots N_4 } \epsilon
 F_{N_1\dots N_4 M}
 -{1\over 96} [\Gamma_{M}{}^{N_1N_2N_3}
(G-G^*)_{N_1N_2N_3}
\cr
-9 \Gamma^{N_1N_2} (G-G^*)_{MN_1N_2}] \epsilon=0~.
\eea
Similarly, taking the difference of (\ref{kseqnb}) and (\ref{kseqnbc}), we find
\bea
(P-P^*)_M \Gamma^M \epsilon+ {1\over 24} (G-G^*)_{N_1N_2N_3}
\Gamma^{N_1N_2N_3} \epsilon=0~.
\eea
It appears that the Majorana condition on the spinor imposes some reality
restrictions on the complex field strengths $P$ and $G$.

\newsection {$SU(4)\ltimes\bR^8$-invariant Killing spinors }

\subsection{The conditions}\la{concon}

We have shown that the canonical form of the most general
IIB $SU(4)\ltimes\bR^8$ invariant
Killing spinor is
$\epsilon= (f-g_2+ig_1) 1+ (f+g_2+i g_1) e_{1234}$,
where  $f,g_1,g_2$ are real functions.
Using the property that $\Gamma^-\epsilon=0$,
one can show that
\be
\e^{w \Gamma^{-+}}\, \epsilon= e^{w}\,\epsilon
\la{gaugefree}
\ee
for any spacetime function $w$. This gauge freedom can be used
to normalize the overall scale of $\epsilon$. We shall use this freedom
in the description of geometry of supersymmetric backgrounds.

To derive the linear system which does not involve gamma matrices
from the Killing spinor equations, we substitute the above spinor $\epsilon$
into the Killing spinor
equations. Then we decompose the vector $SO(9,1)$ representation under $SU(4)$. This
is equivalent to decomposing the frame indices as $A=(+,-,\alpha, \bar\alpha)$.
Consequently, the fluxes and geometry decompose into  $SU(4)$
representations, i.e. $P_{A}$ decomposes as $P_{+}, P_{-}, P_{\a}$ and $P_{\bar\a}$
and similarly for the other fluxes and geometry\footnote{
If the fluxes are complex, like $P$ and $G$, then their various components
do not satisfy the `naive' complex conjugate relations, i.e. $(P_\a)^*\not= P_{\bar\a}$
and similarly for $G$.}.
We also decompose the Killing spinor equations under $SU(4)$ representations using
the decomposition of the fluxes and geometry that we have mentioned and
the decomposition the gamma matrices as
$\Gamma^A=(\Gamma^+, \Gamma^-, \Gamma^\a, \Gamma^{\bar\a})$.
Then we use the properties  $\Gamma^\a 1=\Gamma^- 1=0$ and $\Gamma^{\bar\a} e_{1234}=
\Gamma^-e_{1234}=0$, which we have explained in appendix A,
to rewrite the Killing spinor
equations
 in  the ({\ref{hbasisa}}) basis. Setting every component of the Killing spinor equations
 in this basis to zero,
we derive a linear system
for the fluxes, the geometry, as represented by the Levi-Civita connection of spacetime,
 and the first derivatives of the functions $f, g_1$ and $g_2$.
 Here, we shall present the linear system that
arises from the Killing spinor equations. This system is solved in appendix B.

First, we substitute this spinor into the  (algebraic) Killing spinor equation
({\ref{kseqnb}}), expand the resulting expression in the basis ({\ref{hbasisa}}) and
set every component in this basis
to zero. We  find that ({\ref{kseqnb}}) implies the conditions
\bea
(f+g_2 -i g_1) P_{\bar{\alpha}} +{1 \over 4} (f-g_2 +i g_1) G_{-+\bar{\alpha}}
+{1 \over 4} (f-g_2+ig_1)G_{\bar{\alpha}\b}{}^\b
\cr
+{1 \over 12}(f+g_2+ig_1) \epsilon_{\bar{\alpha}}{}^{\b_1 \b_2 \b_3} G_{\b_1\b_2\b_3} =0~,
\la{done}
\eea
\bea
(f-g_2 -i g_1) P_\alpha +{1 \over 4} (f+g_2+ig_1) G_{-+\alpha}
-{1 \over 4} (f+g_2+ig_1)G_{\alpha \b}{}^\b
\cr
+{1 \over 12} (f-g_2+ig_1) \epsilon_\alpha{}^{{\bar{\b_1}} {\bar{\b_2}} {\bar{\b_3}}}
G_{{\bar{\b_1}} {\bar{\b_2}} {\bar{\b_3}}} =0~,
\la{dtwo}
\eea
\bea
(f+g_2 -i g_1)P_+ +{1 \over 4} (f-g_2 +i g_1) G_{+ \alpha}{}^\a =0~,
\la{dthree}
\eea
\bea
(f-g_2 -i g_1)P_+ -{1 \over 4} (f+g_2 +i g_1) G_{+ \alpha}{}^\a =0~,
\la{dfour}
\eea
and
\be
(f-g_2+ig_1) G_{+{\bar{\alpha}} {\bar{\beta}}}-{1 \over 2} (f+g_2+ig_1)
\epsilon_{{\bar{\alpha}}
{\bar{\beta}}}{}^{\g \d} G_{+ \g \d}=0~.
\la{dfive}
\ee
It is clear that this is a linear system for the fluxes of IIB supergravity and does not
involve gamma matrices.

Next we turn into the Killing spinor equation associated
with the supercovariant derivative (\ref{kseqna}).
In particular the conditions along
the $\a$-frame derivative of the supercovariant connection are
\bea
[D_\a +{1\over2}\Omega_{\a,\b}{}^\b+
{1\over2} \Omega_{\a,-+}
+{i\over4} F_{\a\b}{}^\b{}_{\g}{}^\g
+{i\over2} F_{\a-+\b}{}^\b](f-g_2+ig_1)
\cr
+(f+g_2-ig_1) [{1\over4} G_{\a\b}{}^\b
+{1\over4} G_{-+\a}]=0~,
\la{aone}
\eea
\bea
 (f-g_2+ig_1) [\Omega_{\a,\bar\b_1\bar\b_2}+i F_{\a\bar\b_1\bar\b_2\g}{}^\g
  +i F_{\a -+\bar\b_1\bar\b_2}]
 \cr
 +
(f+g_2-ig_1) [{1\over2} G_{\a\bar\b_1\bar\b_2}
-{1\over4} g_{\a[\bar\b_1} G_{\bar\b_2]\g}{}^\g
-{1\over4} g_{\a[\bar\b_1} G_{\bar\b_2]-+}]
\cr
-
(f+g_2+ig_1) [{1\over2} \Omega_{\a,\g_1\g_2}-{i\over2}
F_{\a\g_1\g_2\d}{}^\d+{i\over2} F_{\a-+\g_1\g_2}
 ] \epsilon^{\g_1\g_2}{}_{\bar\b_1\bar\b_2}
 \cr
 -{1\over8}(f-g_2-ig_1) G_{\a\g_1\g_2} \epsilon^{\g_1\g_2}{}_{\bar\b_1\bar\b_2}=0~,
 \la{atwo}
 \eea
 \bea
[ D_\a -{1\over2} \Omega_{\a,\b}{}^\b+{1\over2} \Omega_{\a,-+}
+{i\over4} F_{\a\b}{}^\b{}_{\g}{}^\g-{i\over2}
F_{\a-+\g}{}^\g](f+g_2+ig_1)
\cr
+ [-{1\over8} G_{\a\g}{}^\g+{1\over8} G_{-+\a}] (f-g_2-ig_1)
\cr
+{i\over12} F_{\a\bar\b_1\bar\b_2\bar\b_3\bar\b_4}
 \epsilon^{\bar\b_1\bar\b_2\bar\b_3\bar\b_4}
(f-g_2+ig_1)-{1\over24}
 \epsilon_{\a}{}^{\bar\b_1\bar\b_2\bar\b_3}
  G_{\bar\b_1\bar\b_2\bar\b_3}(f+g_2-ig_1)=0~,
  \la{athree}
\eea
\bea
[{1\over2} \Omega_{\a,+\bar\b}+{i\over2} F_{\a+\bar\b\g}{}^\g] (f-g_2+i g_1)
+[{1\over16}g_{\a\bar\b} G_{+\g}{}^\g-
{1\over4} G_{+\a\bar\b}](f+g_2-i g_1)
\cr
-{i\over 6} F_{\a+\b_1\b_2\b_3} \epsilon^{\b_1\b_2\b_3}{}_{\bar\b}
(f+g_2+ig_1)=0~,
\la{afour}
\eea
\bea
{i\over12} F_{\a+\bar\b_1\bar\b_2\bar\b_3} (f-g_2+ig_1)
+{1\over32} g_{\a[\bar\b_1} G_{\bar\b_2\bar\b_3]+}
(f+g_2-ig_1)
+{1\over12} (f+g_2+ig_1)
\cr
 [{1\over2} \Omega_{\a,+\g}
-{i\over2} F_{\a+\g\d}{}^\d] \epsilon^\g{}_{\bar\b_1\bar\b_2\bar\b_3} -
{1\over 96}  (f-g_2-ig_1) G_{+\a\g} \epsilon^\g{}_{\bar\b_1\bar\b_2\bar\b_3}=0~.
\la{afive}
\eea

The conditions along the $\bar\a$-frame derivative of the supercovariant connection are
\bea
[D_{\bar\a}+{1\over2} \Omega_{\bar\a,\b}{}^\b+{1\over2} \Omega_{\bar\a, -+}
+{i\over4} F_{\bar\a\b}{}^\b{}_\g{}^\g
+{i\over2} F_{\bar\a-+\b}{}^\b](f-g_2+ig_1)
\cr
+{1\over8} [ G_{\bar\a\b}{}^\b+ G_{\bar\a-+}] (f+g_2-ig_1)
+{i\over12} (f+g_2+ig_1) F_{\bar\a\g_1\g_2\g_2\g_4} \epsilon^{\g_1\g_2\g_3\g_4}
\cr
-{1\over24}
(f-g_2-ig_1)  \epsilon_{\bar\a}{}^{\g_1\g_2\g_3}
G_{\g_1\g_2\g_3}=0~,
\la{abone}
\eea
\bea
[\Omega_{\bar\a, \bar\b_1\bar\b_2}+i F_{\bar\a\bar\b_1\bar\b_2\g}{}^\g
+i F_{\bar\a-+\bar\b_1\bar\b_2}] (f-g_2+ig_1)
+{1\over4} G_{\bar\a\bar\b_1\bar\b_2} (f+g_2-ig_1)
\cr
-(f+g_2+ig_1)[{1\over2}\Omega_{\bar\a,\g_1\g_2}
-{i\over2} F_{\bar\a\g_1\g_2\d}{}^\d+{i\over2} F_{\bar\a-+\g_1\g_2}]
\epsilon^{\g_1\g_2}{}_{\bar\b_1\bar\b_2}
\cr
-(f-g_2-ig_1)[{1\over8} g_{\bar\a\g_1} G_{\g_2\d}{}^\d
+{1\over4}G_{\bar\a\g_1\g_2}-{1\over8} g_{\bar\a\g_1} G_{\g_2-+}]
\epsilon^{\g_1\g_2}{}_{\bar\b_1\bar\b_2}=0~,
\la{abtwo}
\eea
\bea
[D_{\bar\a}-{1\over2} \Omega_{\bar\a,\g}{}^\g+{1\over2} \Omega_{\bar\a,-+}
+{i\over4} F_{\bar\a\g}{}^\g{}_\d{}^\d
-{i\over2} F_{\bar\a-+\g}{}^\g](f+g_2+ig_1)
\cr
+[-{1\over4} G_{\bar\a\g}{}^\g+{1\over4} G_{\bar\a-+}] (f-g_2-ig_1)=0~,
\la{abthree}
\eea
\bea
[{1\over2} \Omega_{\bar\a,+\bar\b}+{i\over2} F_{\bar\a+\bar\b\g}{}^\g] (f-g_2+ig_1)
-{1\over8} G_{+\bar\a\bar\b} (f+g_2-ig_1)
\cr
-{i\over6} (f+g_2+ig_1) F_{\bar\a+\g_1\g_2\g_3} \epsilon^{\g_1\g_2\g_3}{}_{\bar\b}
-{1\over16} (f-g_2-ig_1) G_{\g_1\g_2+} \epsilon^{\g_1\g_2}{}_{\bar\a\bar\b}=0~,
\la{abfour}
\eea
\bea
i F_{\bar\a+\bar\b_1\bar\b_2\bar\b_3} (f-g_2+ig_1)+ (f+g_2+ig_1)
[{1\over2} \Omega_{\bar\a,+\g}
-{i\over2} F_{\bar\a+\g\d}{}^\d] \epsilon^\g{}_{\bar\b_1\bar\b_2\bar\b_3}
\cr
+(f-g_2-ig_1)[-{1\over16} g_{\bar\a\g} G_{+\d}{}^\d
+{1\over4} G_{\bar\a+\g}] \epsilon^\g{}_{\bar\b_1\bar\b_2\bar\b_3}=0~.
\la{abfive}
\eea

The conditions along the $-$-frame derivative of the supercovariant connection are
\bea
[D_-+{1\over2} \Omega_{-,\g}{}^\g+{1\over2} \Omega_{-,-+}
+{i\over4} F_{-\g}{}^\g{}_\d{}^\d](f-g_2+ig_1)
+{1\over4} G_{-\g}{}^\g (f+g_2-ig_1)
\cr
+{i\over12} (f+g_2+ig_1) F_{-\g_1\g_2\g_3\g_4} \epsilon^{\g_1\g_2\g_3\g_4}=0~,
\la{mone}
\eea
\bea
[\Omega_{-,\bar\b_1\bar\b_2}+i F_{-\bar\b_1\bar\b_2\g}{}^\g]
(f-g_2+ig_1)+{1\over2} G_{-\bar\b_1\bar\b_2}
(f+g_2-ig_1)
\cr
-(f+g_2+ig_1) [{1\over2} \Omega_{-,\g_1\g_2}
-{i\over2} F_{-\g_1\g_2\d}{}^\d] \epsilon^{\g_1\g_2}{}_{\bar\b_1\bar\b_2}
\cr
-{1\over4} (f-g_2-ig_1) G_{-\g_1\g_2} \epsilon^{\g_1\g_2}{}_{\bar\b_1\bar\b_2}=0~,
\la{mtwo}
\eea
\bea
[D_--{1\over2} \Omega_{-,\g}{}^\g+{1\over2}\Omega_{-,-+}
+{i\over4} F_{-\g}{}^\g{}_\d{}^\d](f+g_2+ig_1)
\cr
-{1\over4} (f-g_2-ig_1) G_{-\g}{}^\g
+{i\over 12} (f-g_2+ig_1) F_{-\bar\b_1\bar\b_2\bar\b_3\bar\b_4}
\epsilon^{\bar\b_1\bar\b_2\bar\b_3\bar\b_4}=0~,
\la{mthree}
\eea
\bea
[{1\over2} \Omega_{-,+\bar\b}+{i\over2} F_{-+\bar\b\g}{}^\g] (f-g_2+ig_1)
+[-{1\over16} G_{\bar\b\g}{}^\g+{3\over16} G_{-+\bar\b}]
(f+g_2-ig_1)
\cr
-{i\over6} (f+g_2+i g_1) F_{-+\g_1\g_2\g_3} \epsilon^{\g_1\g_2\g_3}{}_{\bar\b}
+{1\over 48} (f-g_2-i g_1) G_{\g_1\g_2\g_3} \epsilon^{\g_1\g_2\g_3}{}_{\bar\b}=0~,
\la{mfour}
\eea
\bea
i F_{-+\bar\b_1\bar\b_2\bar\b_3} (f-g_2+ig_1)
-{1\over8} G_{\bar\b_1\bar\b_2\bar\b_3} (f+g_2-ig_1)
\cr
+
[{1\over2} \Omega_{-,+\g}-{i\over2}
F_{-+\g\d}{}^\d] \epsilon^\g{}_{\bar\b_1\bar\b_2\bar\b_3}
(f+g_2+i g_1)
\cr
+ [{1\over16} G_{\g\d}{}^\d +{3\over16} G_{-+\g}]
(f-g_2-i g_1) \epsilon^\g{}_{\bar\b_1\bar\b_2\bar\b_3}=0~.
\la{mfive}
\eea

The conditions along the $+$-frame derivative of the supercovariant connection are
\bea
[D_++{1\over2} \Omega_{+,\g}{}^\g
+{1\over2} \Omega_{+,-+}+{i\over4} F_{+\g}{}^\g{}_\d{}^\d](f-g_2+ig_1)
\cr
+{1\over8} G_{+\g}{}^\g (f+g_2-ig_1)
+{i\over 12} (f+g_2+ig_1) F_{+\g_1\g_2\g_3\g_4}
\epsilon^{\g_1\g_2\g_3\g_4}=0~,
\la{pone}
\eea
\bea
[\Omega_{+,\bar\b_1\bar\b_2}+i F_{+\bar\b_1\bar\b_2\g}{}^\g]
 (f-g_2+ig_1)+{1\over4} (f+g_2-ig_1) G_{+\bar\b_1\bar\b_2}
\cr
-[{1\over2} \Omega_{+,\g_1\g_2}
-{i\over2} F_{+\g_1\g_2\d}{}^\d] (f+g_2+ig_1) \epsilon^{\g_1\g_2}{}_{\bar\b_1\bar\b_2}
\cr
-{1\over8} (f-g_2-ig_1) G_{+\g_1\g_2} \epsilon^{\g_1\g_2}{}_{\bar\b_1\bar\b_2}=0~,
\la{ptwo}
\eea
\bea
[D_+-{1\over2} \Omega_{+,\g}{}^\g+{1\over2} \Omega_{+,-+}
+{i\over4} F_{+\g}{}^\g{}_\d{}^\d] (f+g_2+ig_1)
\cr
-{1\over8} G_{+\g}{}^\g (f-g_2-ig_1)
+{i\over12} (f-g_2+ig_1)
 F_{+\bar\b_1\bar\b_2\bar\b_3\bar\b_4} \epsilon^{\bar\b_1\bar\b_2\bar\b_3\bar\b_4}=0~,
 \la{pthree}
 \eea
 and
 \bea
 \Omega_{+,+\a}=\Omega_{+,+\bar\a}=0~.
 \la{pfour}
 \eea
As we have already mentioned all the equations that arise from the
Killing spinor equations are linear in the fluxes, geometry and the first derivatives
of the functions $f, g_1$ and $g_2$. Although, this linear system may
appear rather involved
it can be solved to express the  fluxes in terms of the geometry and to find the conditions
on the geometry of the spacetime required by the existence of an $SU(4)\ltimes \bR^8$ invariant
Killing spinor. We remark that we have not used the self-duality condition on the five-form
flux $F$ in  the above conditions. However, the self-duality condition
 will be implemented
in the solution of the linear system in the appendices.

\subsection{Geometry of spacetime}

\subsubsection{Conditions on the geometry}

The expressions of the fluxes in terms of the geometry that solve
the linear system of the previous section and the self-duality
condition for $F$ can be found in appendix B.
Here, we summarize the conditions we have found on the geometry
of the spacetime for backgrounds that admit an $SU(4)\ltimes \bR^8$ invariant
Killing spinor. In this case, it is assumed that the functions $f, g_1, g_2$
of the Killing spinor are generic, i.e. that they do not satisfy any relations.
The conditions on the geometry that arise from the solution of the Killing spinor
equations are
\bea
\Omega_{+,+\a}=\Omega_{\a,+}{}^\a=-\Omega_{\bar\a,+}{}^{\bar\a}=
\Omega_{+,\a}{}^\a+{i g_2\over f} Q_+=0~,
\la{gsuone}
\eea
\bea
\Omega_{\a,+\b}=0~,~~~
\Omega_{\a,+\bar\b}+\Omega_{\bar\b,+\a}=0~,~~~
\la{gsutwo}
\eea
\bea
2\partial_+f+\Omega_{+,-+} f+Q_+ g_1&=&0~,~~~2\partial_+g_1
+\Omega_{+,-+} g_1-{f^2-g_2^2\over
f} Q_+=0~,~~~
\cr
2\partial_+g_2+\Omega_{+,-+} g_2-{g_1 g_2\over f} Q_+&=&0~,~~~
\la{gsuthree}
\eea
\bea
\partial_- (f^2+g_2^2+g_1^2)+\Omega_{-,-+} (f^2+g_2^2+g_1^2)=0~,
\la{gsufour}
\eea
\bea
\partial_{\bar\a} (f^2+g_1^2+g_2^2)+(\Omega_{\bar\a,-+}+\Omega_{-,\bar\a+})
(f^2+g_1^2+g_2^2)=0~,
\la{gsufive}
\eea
\bea
\Omega_{+,\bar\a\bar\b}=0~.
\la{gsusix}
\eea
Observe that the conditions (\ref{gsuthree}) imply that
\bea
\partial_+(f^2+g_1^2+g_2^2)+\Omega_{+,-+} (f^2+g_1^2+g_2^2)=0~.
\la{gsutwob}
\eea
This condition is used later to show that the spacetime admits a Killing vector field.
Observe also that the first equality in the second equation in (\ref{gsuone}) is not
independent but  follows from the second equation in (\ref{gsutwo}). The last equation
(\ref{gsusix}) relates two components of the Levi-Civita connection. In what follows,
we shall focus on the conditions (\ref{gsuone})-(\ref{gsufive}).
Note that $P$ determines the scalars up to a $U(1)$ gauge choice.
So  $Q$ is specified up to a $U(1)$
gauge transformation. Therefore one of the equations in (\ref{gsuthree}) can be used to
determine $Q_+$. However, one can also view them as conditions on the geometry.

\subsubsection{Spacetime forms and the geometry of spacetime }

One way to analyze  the conditions (\ref{gsuone})-(\ref{gsufive})
is to find the spacetime forms that arise from spinor bi-linears and
are associated to the Killing spinor $\epsilon$. Unlike the case of eleven-dimensional
supergravity, the Killing spinor
$\epsilon={1\over \sqrt{2}} [(f-g_2+ig_1) 1+ (f+g_2+i g_1) e_{1234}]$
is complex\footnote{We have normalized the Killing spinor with an
additional factor of $1/\sqrt{2}$.}. Because of this, certain $Pin(9,1)$
and $Spin(9,1)$ invariant
operators\footnote{Such operators are $L_{\pm}$ of appendix A.}
 act non-trivially on $\epsilon$ and give other spinors on the spacetime
 independent from $\epsilon$.
It turns out that it suffices\footnote{Because the spinor $\epsilon$
is Weyl the other operator does not give a new independent spinor.} to consider the spinor
 \bea
 \tilde\epsilon
=C^*\epsilon={1\over \sqrt{2}} [(f+g_2-i g_1) 1+(f-g_2-ig_1) e_{1234}]~.
\eea
The spinor $\tilde \epsilon$ may  {\it not} be a Killing spinor.
 The spacetime
forms bi-linears associated with the pairs $(\epsilon, \epsilon),
(\epsilon, \tilde \epsilon)$ and  $(\tilde\epsilon, \tilde \epsilon)$
can be easily computed using the results of appendix A. In particular,
we find three one-forms
\bea
\kappa(\epsilon, \epsilon)&=&[(f+ig_1)^2- g_2^2] (e^0-e^5)~,~~
\kappa(\tilde\epsilon, \tilde\epsilon)=[(f-ig_1)^2- g_2^2] (e^0-e^5)~,
\cr
\kappa(\epsilon, \tilde\epsilon)&=&[f^2+g_1^2+ g_2^2] (e^0-e^5)~,
\la{suoform}
\eea
a three-form
\bea
\xi(\epsilon, \tilde\epsilon)=2i f g_2 (e^0-e^5)\wedge\omega~,
\eea
and three five-forms
\bea
\tau(\epsilon, \epsilon)&=&{1\over2} (f-g_2+ig_1)^2 (e^0-e^5)\wedge \chi+
{1\over2} (f+g_2+ig_1)^2 (e^0-e^5)\wedge \chi^*
\cr
&&-{1\over2}[(f+ig_1)^2-g^2_2]
(e^0-e^5)\wedge \omega\wedge\omega~,
\cr
\tau(\tilde\epsilon, \tilde\epsilon)&=&{1\over2} (f+g_2-ig_1)^2 (e^0-e^5)\wedge \chi+
{1\over2} (f-g_2-ig_1)^2 (e^0-e^5)\wedge \chi^*
\cr
&&-{1\over2}[(f-ig_1)^2-g^2_2]
(e^0-e^5)\wedge \omega\wedge\omega~,
\cr
\tau(\epsilon,\tilde\epsilon)&=&(f^2-g_2^2+g_1^2) (e^0-e^5)\wedge{\rm Re\,}\chi
-2g_1g_2 (e^0-e^5)\wedge{\rm Im\,}\chi
\cr
&&-{1\over2} [f^2+g_1^2+g_2^2]
(e^0-e^5)\wedge\omega\wedge\omega~.
\eea
Observe that the one-forms are along the same direction but this is not the case
for the five-forms. The three-form vanishes if $g_2=0$ and the $SU(4)$ structure
enhances to a $Spin(7)$ structure. The relations between the various forms
are apparent from their formulae. For example taking the inner product of the three-form
with respect to one of the one-forms, one gets a two-form on the spacetime which
after an appropriate normalization can be interpreted as the K\"ahler form of
an eight-dimensional subspace.

The spacetime  Killing spinor form bi-linears above of  supersymmetric IIB
backgrounds
are complex, unlike those of eleven-dimensional supergravity which are real.
This is not surprising
  because both $\epsilon$ and $\tilde\epsilon$ are complex. It appears that
the geometry of
supersymmetric IIB backgrounds is complex. So  the interpretation
of the various geometric condition that arise from supersymmetry for IIB backgrounds
 may require
to complexify the  tangent bundle and the bundle
of forms of the spacetime.

To investigate further the geometry of spacetime,
we introduce a frame such that
\bea
ds^2=2 e^+ e^-+ 2\delta_{\a\bar\b} e^\a e^{\bar\b}~,
\la{metrherm}
\eea
where $e^-=1/\sqrt{2}\, (-e^0+e^5)$, $e^+=1/\sqrt{2}\, (e^0+e^5)$ and $e^\a, e^{\bar\a}$
is a Hermitian frame.
After some rescaling, the third one-form in (\ref{suoform}) can be written as
\bea
\kappa=-{1\over\sqrt 2} \kappa(\epsilon, \tilde\epsilon)=
(f^2+g_1^2+g_2^2) e^-~.
\eea
We denote the associated null vector field by $X$, i.e. $X=(f^2+g_1^2+g_2^2) e_+$,
where $e^A(e_B)=\delta^A_B$ and $e_B$ is the coframe.
Using  the first equation in (\ref{gsuone}) and (\ref{gsutwob}), one can show that
$X$ is self-parallel, i.e. it satisfies the equation
\bea
X^B \nabla_B X^A=0~.
\eea
In addition $X$ is Killing, $\nabla_A X_B+\nabla_B X_A=0$. This follows from the
equations (\ref{gsuthree})-(\ref{gsufive}) and (\ref{gsutwob}). We have not found
an interpretation for all the  conditions in (\ref{gsuone}). But some of them
imply
\bea
X^A \nabla_A \chi_{B_1\dots B_4} (\chi^*)^{B_1\dots B_4}=0~.
\eea
The $SU(4)$ structure on the spacetime is generic. If there were a restriction
on it, there must have been a relation in (\ref{gsuone})-(\ref{gsusix})
 that restricted the components
of the Levi-Civita connection $\Omega$ along the $SU(4)$ directions
 $e^\a, e^{\bar\a}$.  But there
is no such relation.

One can introduce coordinates adapted to the null, Killing vector field $X$, $
X={\partial\over\partial u}$
and write the spacetime metric as
\bea
ds^2=2 (W dv+m_I dy^I) ( U du+ V dv+  n_I dy^I )+ \gamma_{IJ} dy^I dy^J~,
\la{coormetr}
\eea
where $W, U, V, m_I, n_I$ and $\gamma_{IJ}$ are functions of all coordinates,
$I,J=1,\dots,8$.
Since $X$ is Killing, all the components of the metric are independent of $u$.
In addition $U=f^2+g_1^2+g_2^2$. To see this, we introduce the frame
\bea
e^-=W dv+m_I dy^I~,~~~ e^+=U du+ V dv+  n_I dy^I~,~~~e^i= e^i_J dy^J~,
\la{coorframe}
\eea
where $\gamma_{IJ}=\delta_{ij}  e^i_I e^j_J dy^I dy^J$.
Then we have
\be
X=(f^2+g_1^2+g_2^2)e_+={\partial\over\partial u}~,
\ee
where $e_B$ is
\bea
e_+= U^{-1} {\partial\over\partial u}~,~~~ e_-= W^{-1} {\partial \over\partial v}-
(U W)^{-1} V{\partial \over\partial u}~,~~~
\cr
e_i= e_i^J {\partial\over\partial y^J}+(-{n_i\over U}+{V m_i\over W U})
{\partial\over\partial u}
- {m_i\over W} {\partial\over\partial v}~,
\la{coorcoframe}
\eea
where $e^i_I e^I_j=\delta^i{}_j$, $m_i= m_I e^I_i$ and $n_i= n_I e^I_i$.
However, the Killing vector field can be written as
\bea
X=(f^2+g_1^2+g_2^2)e_+=(f^2+g_1^2+g_2^2)U^{-1}
{\partial\over\partial u}={\partial\over\partial u}~.
\eea
Therefore $U=f^2+g_1^2+g_2^2$.

To write the metric in (\ref{coormetr}), we have separated a coordinate $v$.
Unlike $u$,  there is no natural way to choose the coordinate $v$,
i.e. the conditions on the geometry
(\ref{gsuone})-(\ref{gsusix}) implied by the Killing spinor equations do not
lead to a definition of $v$. Because of this, there is no natural way to
define an eight-dimensional submanifold $\Sigma$ in the spacetime which
can be identified as that that has an $SU(4)$ structure.

A simplification of the conditions on the geometry (\ref{gsufour}) and (\ref{gsufive}),
and of the local expression of the metric can be made by fixing a gauge for the
transformation (\ref{gaugefree}) that scales the Killing spinor $\epsilon$
with a positive
 spacetime function\footnote{Further gauge fixing is possible. For example, one can use
 $R_{++,AB}=0$ to set $\Omega_{+,AB}=0$.}. For example, one can fix this gauge freedom by setting
 \bea
 U=f^2+g_1^2+g_2^2=1~.
 \eea
 In this gauge (\ref{gsufour}) and (\ref{gsufive}) imply that
 $\Omega_{-,-+}=0$ and $\Omega_{\bar\a, -+}=-\Omega_{-, \bar\a+}$.
 In addition, one can set $U=1$ in (\ref{coormetr}), (\ref{coorframe}) and
  (\ref{coorcoframe}). In turn, (\ref{gsuthree}) implies
  that $Q_+=-2 g_1^{-1} \partial_+f$
  and $\partial_+ (f g_2)=0$.

Examples of spacetimes that have the structure we have investigated in this section
are  Lorentzian extensions of  one-parameter
families of a manifolds with a generic $SU(4)$ structure. The metric
on such a family can be written as
\bea
d\hat s^2= 2 B^2 dv^2+ \hat\gamma_{IJ} (dy^I+ A^I dv)  (dy^J+ A^J dv)~,
\la{fammetr}
\eea
where $B$, $\gamma_{IJ}$ and  $A$ depend on the coordinates
 $v, y^I$, $I=1,\dots,8$. The component $A$
can be thought of as a non-linear connection of the family.
Setting $u={\rm const}$ in (\ref{coormetr}), we find that
\bea
d\tilde s^2&=&2 (W dv+m_I dy^I) ( V dv+  n_I dy^I )+ \gamma_{IJ} dy^I dy^J
\cr
 &=&2 UV dv^2+ 2 (Wn_I+V m_I) dv dy^I+ (\gamma_{IJ} +m_I n_J) dy^I dy^J
 \eea
Comparing this with (\ref{fammetr}), we get
\bea
UV=B^2+{1\over 2} \hat\gamma_{IJ} A^I A^J
\cr
Wn_I+V m_I=\hat\gamma_{IJ} A^J
\cr
\gamma_{IJ} +m_{(I} n_{J)}=\hat\gamma_{IJ}~.
\eea
It is clear that to specify the geometry of spacetime additional
data are needed which determine the extension of the $SU(4)$ family to
a Lorentian manifold. In turn these are related to the type of reduction
of the $Spin(9,1)$ structure to the $SU(4)\ltimes\bR^8$ structure
 of  ten-dimensional spacetime.

\subsection{A null geodesic congruence}

An alternative way to interpret the conditions on the geometry of the spacetime
(\ref{gsuone})-(\ref{gsusix})
is to use  geodesic congruences. Since $X$ is null and self-parallel,
it defines a null geodesic congruence in the spacetime. In addition $X$ is Killing.
This implies that the null geodesic congruence is divergence free,
i.e.
\be
\nabla_A X^A=0~,
\ee
and shear free\footnote{If $X$ is a timelike or spacelike vector,
then the last term in the  formulae below is weighted by the length square
of $X$ such that $s_A{}^A=0$.}, i.e.
\bea
s_{AB}=\nabla_A X_B+\nabla_B X_A-{2\over 9} \theta (g_{AB}- X_A X_B)=0~.
\la{she}
\eea
However, the geodesic congruence is not rotation free, i.e.
\bea
r_{AB}=\nabla_A X_B-\nabla_B X_A
\la{rot}
\eea
does not vanish. Because of this, one cannot introduce Penrose coordinates
on the spacetime \cite{penrose} along the null geodesic
congruence defined by $X$. In particular, this implies
that one cannot define a coordinate $v$. Nevertheless, if one choose a null
geodesic in the null geodesic congruence defined by $X$, it is always possible
to embed it in a rotation free null geodesic congruence and then
take the associated Penrose limit. It turns out that the plane wave at the limit
is Minkowski space because $X$ is Killing and so the plane wave metric in Rosen
coordinates does not depend on the affine parameter of the geodesic, see \cite{blau}.

\newsection {$Spin(7)\ltimes\bR^8$-invariant Killing spinors }

\subsection{The conditions on the geometry}

The most general IIB Killing spinor
 which is invariant under $Spin(7)\ltimes\bR^8$ is
$\epsilon= (f+ig) (1+ e_{1234})$, where $f,g$ are  real functions .
It  is always possible to choose a gauge  such that
$\epsilon=(1+ e_{1234})$.  This  is because the the Killing spinor equations
of IIB supergravity
are covariant under the
 transformation
$$e^{w\Gamma^{-+}+i\varphi}$$
 after an appropriate rotation of $G$. The parameters $w$ and $\varphi$
 can be chosen such that
 $g=0$ and  $f=1$.
However for many computations it is useful to have the freedom to
rotate with $e^{i\varphi}$. Because of this we shall allow $f, g$ to
be arbitrary. However as we shall explain later the
$e^{w\Gamma^{-+}}$ transformation can be used to simplify the
conditions on the geometry of spacetime. One can choose either
$f\not=0$ or $g\not=0$. In what follows we shall take $f\not=0$. The
conditions on the geometry and
 fluxes that arise from the Killing spinor equations for $\epsilon= (f+ig) (1+ e_{1234})$
 can be easily derived from those of an $SU(4)\ltimes\bR^8$
 invariant  spinor  by setting $f=f$, $g_1=g$ and $g_2=0$
in the conditions of section  (\ref{concon}). Because of this,
we shall not state the conditions
 for supersymmetry again. Instead, we summarize the conditions for the geometry.
 The solution of the Killing spinor equations  and the expressions of
 the fluxes in terms of the geometry  are given in appendix C.
 The expressions for the fluxes have been simplified using the self-duality of $F$.

The conditions on the spacetime geometry for backgrounds
that admit the  $Spin(7)\ltimes \bR^8$ invariant  Killing  spinor
 $\epsilon= (f+ig) (1+ e_{1234})$  are
\bea
\Omega_{+,\a+}=\Omega_{\a,+}{}^\a=\Omega_{+,\a}{}^\a=0~,
\la{gspone}
\eea
\bea
\Omega_{\a,+\bar\b}+\Omega_{\bar\b,+\a}=0~,~~~\Omega_{\a,+\b}+\Omega_{\b,+\a}=0~,
\la{gsptwo}
\eea
\bea
2\partial_+ f+\Omega_{+,-+} f+ Q_+ g=0~,~~~
2\partial_+ g +\Omega_{+,-+} g- Q_+ f=0~,
\la{gspthree}
\eea
\bea
\partial_-(f^2+g^2) +\Omega_{-,-+} (f^2+g^2)=0~,
\la{gspfour}
\eea
\bea
\partial_{\bar\a} (f^2+g^2)+(\Omega_{\bar\a,-+}
+ \Omega_{-,\bar\a +}) (f^2+g^2)=0~,
\la{gspfive}
\eea
\bea
\Omega_{\a,+\b}-{1\over2} \epsilon_{\a\b}{}^{\bar\g_1\bar\g_2}
 \Omega_{\bar\g_1,+\bar\g_2}=0~,
\la{gspsix}
\eea
\bea
\Omega_{+,\bar\a\bar\b}-{1\over2} \epsilon_{\bar\a\bar\b}{}^{\g_1\g_2}
\Omega_{+,\g_1\g_2}=0~.
\la{gspseven}
\eea
The conditions that we have found for the geometry of supersymmetric
backgrounds with a $Spin(7)\ltimes \bR^8$ invariant Killing spinor
are closely related to those that we have derived for backgrounds
with an $SU(4)\ltimes \bR^8$ invariant Killing spinor. However there
are some differences. For example the second equation in (\ref{gsptwo})
is different from the first equation in (\ref{gsutwo}). The rest of
the equations in (\ref{gsptwo})-(\ref{gspfive})
are related to (\ref{gsutwo})-(\ref{gsufive}) after setting $g_2=0$.
In addition  (\ref{gsuone}) is same as
in (\ref{gspone}) after setting $g_2=0$. The equations
(\ref{gspsix}) and (\ref{gspseven})  differ from (\ref{gsusix}) and the
second equation in (\ref{gsptwo}).

\subsection{The geometry of spacetime}

To interpret some of the geometric conditions (\ref{gspone})-(\ref{gspseven})
that we have derived from the Killing spinor equations, we shall use the
spacetime form bi-linears.
The $Spin(7)\ltimes\bR^8$ invariant Killing  spinor\footnote{We have normalized
the spinor with an additional factor of $1/\sqrt{2}$.}
$\epsilon={1\over\sqrt{2}}(f+ig) (1+ e_{1234})$ is complex. Because of this,
as we have explained in the case of an $SU(4)\ltimes\bR^8$ invariant Killing spinor,
we should consider the spacetime forms associated with the pairs
$(\epsilon, \epsilon)$, $(\epsilon, \tilde\epsilon)$
and $(\tilde\epsilon, \tilde\epsilon)$,
where $\tilde\epsilon=C * \epsilon={1\over\sqrt{2}}(f-ig) (1+ e_{1234})$
and $C$ is the charge conjugation matrix, see appendix A.
We can easily compute these spacetime forms using the formulae that we have collected
in appendix A. In particular, we find three one-forms
\bea
\kappa(\epsilon, \epsilon)&=&(f+ig)^2 (e^0-e^5)~,~~~
\kappa(\tilde\epsilon, \tilde\epsilon)=(f-ig)^2 (e^0-e^5)~,~~~
\cr
\kappa(\epsilon, \tilde\epsilon)&=&(f^2+g^2) (e^0-e^5)~,
\eea
and three five-forms
\bea
\tau(\epsilon, \epsilon)&=& (f+ig)^2 (e^0-e^5)\wedge \phi~,~~~
\tau(\tilde\epsilon, \tilde\epsilon)= (f-ig)^2 (e^0-e^5)\wedge \phi~,
\cr
\tau(\epsilon, \tilde\epsilon)&=& (f^2+g^2) (e^0-e^5)\wedge \phi~,
\eea
where
\bea
\phi={\rm Re} \chi-{1\over2} \omega\wedge\omega~.
\eea
It is easy to recognize from the expressions  for $\chi$ and $\omega$ in appendix A
that $\phi$ is the usual $Spin(7)$-invariant four-form. Some of the
one-forms and five-forms above are complex as  may have been expected
because the spinors $\epsilon$ and $\tilde \epsilon$ are complex.
Unlike the $SU(4)\ltimes \bR^8$ invariant case,
 the three-form bi-linears
vanish. This may have been expected because there is no apparent
 $Spin(7)$ invariant three-form which can be constructed on a
 ten-dimensional
 manifold.

Writing the metric as (\ref{metrherm}) by introducing the frame
$e^+, e^-, e^\a, e^{\bar\a}$,
we  can write the third one-form bi-linear  as
$\kappa=-{1\over\sqrt {2}}\kappa(\epsilon, \tilde\epsilon)= (f^2+g^2) e^-$.
The associated vector field  is  $X=(f^2+g^2) e_+$,
where $e^A(e_B)=\delta^A{}_B$. As in the $SU(4)\ltimes \bR^8$ case,
the first condition in (\ref{gspone}) and the conditions
(\ref{gsptwo})-(\ref{gspfive}) imply
that the null vector field $X$ is {\it self-parallel} and {\it Killing}.
Therefore $X$ defines a divergence and shear free null geodesic congruence.
One can adapt coordinates along $X$ and write the spacetime metric as
in (\ref{coormetr}) with $U=f^2+g^2$, see \cite{gutowski} for a similar
coordinate system on the spacetime of eleven-dimensional backgrounds
with a null Killing spinor.
The construction of a local expression for the metric
and the introduction of a local frame on the spacetime
 can be done as
in the $SU(4)\ltimes \bR^8$ case that we have presented in section three.
Because of this, we shall not repeat the construction here.
The $Spin(7)$ structure on the spacetime is generic. There are no relations
between the components of the metric that lie along the $(e^\a, e^{\bar\a})$
directions.

One can use the transformation that scales the Killing spinor $\epsilon$
with a positive  spacetime function, as in (\ref{gaugefree}), to set $f^2+g^2=1$.
Then (\ref{gspfour}) and (\ref{gspfive}) imply that $\Omega_{-,-+}=0$
and $\Omega_{-,\a+}=-\Omega_{\a,-+}$. In adition, (\ref{gspthree})
gives $Q_+=- 2g^{-1} \partial_+ f$.

It remains to interpret the second condition in (\ref{gspone}), and
the conditions (\ref{gspsix}) and (\ref{gspseven}). These can be
combined. In particular they imply that  $\Omega_{+,ij}$ and
$\Omega_{[i,j]+}$ lie in the Lie algebra of $Spin(7)$, or
equivalently (\ref{gspone}) (\ref{gspsix}) and (\ref{gspseven})
imply that
\bea
 \Omega_{+,ij}|_{\Lambda_{7}^2(\bR^8)}=\Omega_{[i,j]+}|_{\Lambda_{7}^2(\bR^8)}=0~,
\eea
where we have used the
decomposition of $\Lambda^2(\bR^8)=spin(7)\oplus \Lambda_{7}^2(\bR^8)$
under $Spin(7)$.

Examples of  spacetimes with the above structure
are Lorentzian extensions of  one-parameter families
of  eight-dimensional manifolds with  generic $Spin(7)$ structures.
The relation between the metric on the family and  that of spacetime
can be described  as for the $SU(4)\ltimes\bR^8$ case which has been
presented in section three.

\newsection{Pure spinors as Killing spinors}\la{pcondition}

\subsection{Conditions on the geometry}

A special class of supersymmetric backgrounds with an $SU(4)\ltimes\bR^8$
invariant Killing spinor is that  for which the Killing spinor is in addition
pure. As we have mentioned in the introduction, there are different definitions
of what a pure spinor is. Here we adopt the definition that a pure spinor
is that for which the associated one-form bi-linear vanishes.
Applying this definition to the $SU(4)\ltimes\bR^8$ invariant spinor
$\eta=a 1+ b e_{1234}$, $a,b\in \bC$, it can be shown using the results
in appendix A, that $\eta$ is pure iff $a b=0$.
In particular for the $SU(4)\ltimes\bR^8$ invariant Killing  spinor $\epsilon$, we find
\be
(f+i g_1)^2-g_2^2=0~,
\ee
which in turn gives
\bea
g_1=0~,~~~~~f^2=g_2^2~,
\eea
i.e. $g_1=0$ and $f=\pm g_2$.

First consider the case, $g_1=0$ and  $f=-g_2$.
The pure spinor\footnote{Observe that the spinor $1$ is annihilated
by half of the gamma matrices
so it is a pure spinor according to the other
 definition that  has been mentioned
in the introduction.} is
\bea
\epsilon= h\, 1~,
\eea
where $h=2f$.
To derive the conditions required for $\epsilon= h\, 1$ to be a Killing spinor,
we simply set $g_1=0$ and $f=-g_2$ in the conditions of section (\ref{concon})
that we have derived for the existence of an $SU(4)\ltimes\bR^8$
invariant Killing spinor. Because of this, we shall not
state these conditions again.  Instead, we summarize
the conditions on the geometry that are required for $ h\,1$ to be
a Killing spinor. The derivation of these formulae can be found in appendix D.

The conditions on the geometry of a background to admit $\epsilon= h\, 1$
as Killing spinor are
\bea
\Omega_{+,+\a}=
\Omega_{\a,+}{}^\a=-\Omega_{\bar\a,+}{}^{\bar\a}=-\Omega_{+,\a}{}^\a+i Q_+=0~,
\la{gpsuone}
\eea
\bea
\Omega_{\a,+\bar\b}+\Omega_{\bar\b,+\a}=0~,~~~~\Omega_{\a,+\b}=0~,
\la{gpsutwo}
\eea
\bea
2\partial_+\log h+\Omega_{+,-+}&=&0~,~~~ 2\partial_-\log h+\Omega_{-,-+}=0~,
\cr
2\partial_{\bar\a}\log h&+& \Omega_{\bar\a,-+}+\Omega_{-,\bar\a+}=0~,
\la{gpsuthree}
\eea
\bea
\Omega_{+,\a\b}=0~,~~~
\la{gpsufour}
\eea
\bea
i Q_{\bar\a}-2\Omega_{-,+\bar\a}-\Omega_{\bar\a,\b}{}^\b+\Omega_{\b,\bar\a}{}^\b=0~.
\la{gpww}
\eea
As expected these conditions on the geometry resemble those of the generic
$SU(4)\ltimes \bR^8$ invariant spinor in (\ref{gsuone})-(\ref{gsusix}).
The simplification in the above conditions appears because the Killing
spinor $ h\, 1$ depends on one function while the generic $SU(4)\ltimes \bR^8$
invariant spinor depends on three functions. The expressions for the fluxes in terms of the geometry
 are summarized in appendix D. The main difference between the pure spinor case
 and the generic $SU(4)\ltimes \bR^8$ case is the condition (\ref{gpww}).
 This condition can be rewritten as
 \bea
 Q_j J^j{}_i+ 2 \Omega_{-,+i}+ (W_5)_i=0~,
 \eea
where $W_5= \chi\lc d\chi$ is a Gray-Hervella class. However this equation
is valid only if it is restricted along the $e^i$ frame directions as indicated.

Next consider the other pure spinor $k\, e_{1234}$. Again, the conditions
that arise from the Killing spinor equations for $k\, e_{1234}$ to be Killing spinor
can be derived from those we have derived for
case of a generic $SU(4)\ltimes \bR^8$ invariant spinor. For this,
we set $g_1=0$ and $h=2 f=2 g_2$ in  the conditions of section (\ref{concon}).
We shall not present here the derivation of the solution to the resulting equations.
This can be found in appendix E. It turns out that the  conditions
on the geometry of a background to admit $\epsilon= k\, e_{1234}$
as Killing spinor are the same as those given for the spinor $h\, 1$
(\ref{gpsuone})-(\ref{gpsufour}) after replacing the function $h$ with the function
$k$, expect for the sign of $Q_+$ in (\ref{gpsuone}) and in (\ref{gpww}).
The expressions for the fluxes in terms of the geometry
 are given in appendix E.

\subsection{Geometry and spacetime forms}

We examine the geometry of a spacetime with Killing spinor
$h\, 1$. The analysis for the Killing spinor $k\, e_{1234}$ is similar and
it will be omitted. Since $h\, 1$ is a pure spinor,
  the associated one-form bi-linear vanishes. But as we have explained,
if $\epsilon$ is defined on the spacetime, then so is
$\tilde\epsilon=C*(\epsilon)= h e_{1234}$  even though that it may not be  a Killing
spinor. The spacetime form bi-linears associated with the pairs
o $(\epsilon, \epsilon)$, $(\epsilon, \tilde\epsilon)$ and
$(\tilde\epsilon, \tilde\epsilon)$
are
 a one-form
\bea
\kappa(\epsilon, \tilde\epsilon)&=&h^2 (e^0-e^5)
\eea
a three-form
\bea
\xi(\epsilon, \tilde\epsilon)=-i h^2 (e^0-e^5)\wedge\omega~
\eea
and three five-forms
\bea
\tau(\epsilon, \epsilon)&=&h^2 (e^0-e^5)\wedge \chi~,~~~
\tau(\tilde\epsilon, \tilde\epsilon)=h^2(e^0-e^5)\wedge \chi^*~,
\cr
\tau(\epsilon,\tilde\epsilon)&=&-{h^2\over2}
(e^0-e^5)\wedge\omega\wedge\omega~.
\eea
We introduce the frame $e^+, e^-, e^\a, e^{\bar\a}$ and write the metric as in
(\ref{metrherm}).
After some rescaling, the  one-form can be written as
\bea
\kappa=-{1\over\sqrt 2} \kappa(\epsilon, \tilde\epsilon)=
h^2 e^-~.
\la{kxxx}
\eea
The associated null vector field with $X$ is self-parallel and Killing.
The geometry of the spacetime is similar to that described for the
generic $SU(4)\ltimes\bR^8$ invariant spinor. For example the metric
in local coordinates
is given as in (\ref{coormetr}) with $U=h^2$. In the pure spinor case,
one also has the condition (\ref{gpww}). If the spacetime is constructed
as an one parameter family of manifolds, then the $SU(4)$ structure of the
eight-dimensional manifolds is restricted by this condition.

\newsection{Backgrounds with two pure Killing spinors}

\subsection{Geometry and fluxes}

Combining the results we  present in section  five and in  appendices D and E,
we shall give  the conditions on the geometry and the fluxes
implied by the Killing spinor equations for supersymmetric backgrounds
with Killing spinors $\epsilon= h\, 1$ and $\eta= k\, e_{1234}$.
Thus these backgrounds admit two pure $SU(4)\ltimes\bR^8$ invariant
Killing spinors and they constitute a class of supersymmetric backgrounds
with two supersymmetries.

Since the conditions (\ref{gpsuthree}) are the same for both spinors,
it is easy to see that up to an overall constant scale $h=k$. After some
computation,
the conditions on the geometry required for a background to have
Killing spinors $\epsilon= h\, 1$ and $\eta= h\, e_{1234}$ are
\bea
 \Omega_{+,+\a}=\Omega_{+,\a\b}=\Omega_{\a,+\bar\b}=\Omega_{\a,+\b}=
 \Om_{+,\g}{}^\g=\Om_{-,\g}{}^\g=0~,
 \la{ppsuone}
 \eea
 \bea
 2\partial_+\log h+\Omega_{+,-+}=0~,~~~2\partial_-\log h+\Omega_{-,-+}=0~,
 \cr
 2\partial_{\bar\a}\log h+\Omega_{\bar\a,-+}+\Omega_{-,\bar\a+}=0~,
 \eea
 \bea
\Om_{[\g_1,\g_2\g_3]}=0~,~~~ \Om_{\bar\a,\b}{}^\b=-\frac{1}
{3}\Om_{\b,}{}^\b{}_{\bar\a}=\Om_{-,+\bar\a}~,
 \cr
\Om_{\a,\bar\b_1\bar\b_2}-2\Om_{-,+[\bar\b_1}g_{\bar\b_2]\a}=0~.
\la{ppconsuf}
 \eea

The expressions for the $G$ fluxes in terms of the $F$ fluxes and
geometry required by supersymmetry are
\bea
G_{+\a\b}=G_{+\bar\a\bar\b}=G_{-+\a}=G_{-+\bar\a}=G_{\a\b}{}^\b
=G_{\bar\a\b}{}^\b=G_{+\a\bar\b}=0~,
\eea
\bea
G_{-\a\b}&=&\ep_{\a\b}{}^{\bar\g_1\bar\g_2}(\Om_{-,\bar\g_1\bar\g_2}+i
F_{-\bar\g_1\bar\g_2\d}{}^\d)~,~~~
G_{-\bar\a\bar\b}=\ep_{\bar\a\bar\b}{}^{\g_1\g_2}(\Om_{-,\g_1\g_2}-i
F_{-\g_1\g_2\d}{}^\d)~,
\cr
G_{-\g}{}^\g&=&\frac{i}{3}F_{-\bar\b_1\bar\b_2\bar\b_3\bar\b_4}
\ep^{\bar\b_1\bar\b_2\bar\b_3\bar\b_4}~,
\eea
\bea
G_{\a\bar\g_1\bar\g_2}=\Om_{\a,\d_1\d_2}\ep^{\d_1\d_2}{}_{\bar\g_1\bar\g_2}~,~~~
G_{\g_1\g_2\bar\a}=\Om_{\bar\a,\bar\d_1\bar\d_2}\ep^{\bar\d_1\bar\d_2}{}_{\g_1\g_2}~,
\la{ppGaabab}
\eea
\bea
G_{\g_1\g_2\g_3}=4\Om_{-,+\bar\a}\,\ep^{\bar\a}{}_{\g_1\g_2\g_3}~,~~~
G_{\bar\g_1\bar\g_2\bar\g_3}=4\Om_{-,+\a}\,\ep^{\a}{}_{\bar\g_1\bar\g_2\bar\g_3}~.
\la{ppGaaa}
\eea
Similarly, the conditions on the $F$ fluxes are
 \bea
 F_{-+\g_1\g_2\g_3}=F_{+\a\b\g\d}=F_{+\a\bar\b_1\bar\b_2\bar\b_3}
 =F_{+\a\bar\b\g}{}^\g=F_{\a\bar\b_1\bar\b_2\bar\b_3\bar\b_4}=0~,
  \eea
\bea
&F_{-+\bar\a\g}{}^\g=F_{+\a}{}^\a{}_\d{}^\d=F_{-+\a\bar\b_1\bar\b_2}=0~,
~~~ F_{-\g}{}^\g{}_\d{}^\d=2 Q_-~,
\eea
and
\bea
\ep^{\bar\b_1\bar\b_2\bar\b_3\bar\b_4}F_{-\bar\b_1\bar\b_2\bar\b_3\bar\b_4}=-F_{-\b_1\b_2\b_3\b_4}\ep^{\b_1\b_2\b_3\b_4}~.
\eea
Finally, the conditions on the scalars are
\bea
P_+=Q_+=0~,~~~Q_{\a}=0~,~~~ P_\a=-2\Om_{-,+\a},~~~ P_{\bar\a}=-2\Om_{-,+\bar\a}~.
\la{ppPa}
\eea
The components of the fluxes that do not appear in the above equations
are not specified by the Killing spinor equations.

The conditions on the geometry are investigated in the next section.
Since the flux $G$ is complex, the two conditions in
(\ref{ppGaabab}) and in (\ref{ppGaaa}) are independent. In
particular, the conditions (\ref{ppGaabab}) imply that
$G_{\a\bar\b\bar\g}$ is the complex conjugate of  $G_{\bar\a\b\g}$.
In addition, (\ref{ppPa}) implies that $P_\a$ is the complex
 conjugate of $P_{\bar\a}$.

\subsection{Geometry}

The spacetime forms associated with the Killing spinors $\epsilon=f \,1$
and $\eta=h\, e_{1234}$ have also been given in section 5.2. The
vector field $X$ associated to the one-form $\kappa$ (\ref{kxxx}) is a null, self-parallel,
Killing vector as for the other supersymmetric
backgrounds with $SU(4)\ltimes\bR^8$ invariant
Killing spinors. However, some of the conditions
on the geometry in this case are more restrictive.
For example $\Omega_{\a,\b+}=0$ vanishes instead of   $\Omega_{(\a,\b)+}=0$.
Introducing local coordinates for the metric as in (\ref{coormetr}),
the conditions $\Omega_{\a,\b+}=\Omega_{\a,\bar\b +}=0$ in (\ref{ppsuone})
imply that locally
$m_I=\partial_I m$ for some $m=m(v, y)$.

Furthermore, there are restrictions on the   $SU(4)$ structure
of the ten-dimensional spacetime given in equation (\ref{ppconsuf}).
All the conditions on the geometry we have found can be reexpressed as the vanishing
of certain $SU(4)\ltimes\bR^8$ irreducible representations of the covariant derivative
of the spacetime form  bi-linears of the Killing spinors as in \cite{grayhervella}.
We shall not give an exhaustive list because this is just a rewriting of the conditions
on the geometry we have already expressed in terms of the spin connection $\Omega$.
For example, one can show that $(\nabla\omega)_{[\a\b\g]}=0$ implies
 $\Omega_{[\a,\b\g]}=0$.

\newsection{Conclusions}

We have used the method of \cite{gran} to directly solve the Killing spinor
equations and the self-duality condition on the five-form field strength
 of IIB supergravity for one $Spin(7)\ltimes\bR^8$ or one $SU(4)\ltimes\bR^8$
invariant Killing spinor. In both cases, we have found the conditions
on the geometry of
the spacetime required by supersymmetry. One difference
with similar computations in eleven-dimensions \cite{pakis, gutowski}
 is that in IIB supergravity the Killing spinors
depend on more than one spacetime function. As an example of our construction,
we have presented the solution to the Killing spinor equations for backgrounds
that admit one and two pure $SU(4)\ltimes\bR^8$ invariant Killing spinors.
In this case, the Killing spinor equations simplify. Nevertheless,
 the geometry of
the spacetime  resembles that of  backgrounds with a
generic $SU(4)\ltimes\bR^8$ invariant Killing spinor.
There is another class
of supersymmetric IIB backgrounds with one supersymmetry for which the
Killing spinor is $G_2$ invariant. We shall present
this case elsewhere \cite{gju}.

We have computed the spacetime forms that are associated to
 Killing spinor bi-linears.
It turns out that in general the forms are complex.
This is not surprising because the Killing spinors of IIB supergravity are
complex. Therefore, the spinorial geometry of IIB backgrounds appears
to be associated with complex geometry, i.e. it requires the complexication
of the tangent bundle and of the bundle of forms of the spacetime.
This is unlike the spinorial geometry of
eleven-dimensional backgrounds which is real.
In addition, the spacetime forms in both $Spin(7)\ltimes\bR^8$
and $SU(4)\ltimes\bR^8$
cases are related to
SLAG and Cayley  calibration forms, see e.g. \cite{harvey}.
However, in the context of
supergravity these forms are not closed and so
define generalized calibrations \cite{calone, caltwo}.
It seems that IIB supersymmetric backgrounds admit generalized calibrated cycles.
Such cycles are the supersymmetric solutions of D-brane worldvolume
actions coupled to Ramond-Ramond fields in the absence of $B$ field.

We have seen that the geometry of some supersymmetric IIB backgrounds can be described
using the  reduction of the structure group from $Spin(9,1)$ to $SU(4)\ltimes\bR^8$
and to $Spin(7)\ltimes\bR^8$. It is clear that such reductions
can be understood in a similar way as that proposed
by Gray-Hervella  \cite{grayhervella} for almost Hermitian manifolds
and further developed in \cite{fernandez, salamonb, cabrera}, see also \cite{ivanov}.
However to our knowledge, there is no systematic investigation
of the reduction that occurs in the context of IIB supergravity.
Supersymmetric IIB backgrounds with more than one Killing spinor
may require even more exotic reductions of $Spin(9,1)$.

\section*{Acknowledgements}

The work of U.G.~is funded by the
Swedish Research Council. This work has been partially supported by the
PPARC grant PPA/G/O/2002/00475. J.G. thanks EPSRC for support.

\setcounter{section}{0}

\appendix{Spinors}

The realization of spinors in terms of forms can be found for example
in \cite{lawson, harvey}. This
 has  been used   in \cite{wang} to investigate the parallel spinors and
associated  forms on special holonomy manifolds. The description of spinors
in \cite{lawson, harvey} extends over several chapters and \cite{wang}
 does not describe the construction
for Lorentzian manifolds. Because of this, in this appendix, we summarize
the essential information needed to realize the spinors of $Spin(9,1)$ in terms of
forms.

Let $V=\bR^{9,1}$ be a real vector space equipped
with the  Lorentzian inner product $<\cdot,\cdot>$. Introduce an
orthonormal  basis $e_1, \dots,e_9, e_0$, $e_0$ is along the time direction,
and take the subspace $U=\bR<e_1,\dots,e_5>$ spanned with respect to the
first five basis vectors $e_1,\dots,e_5$.
The space of Dirac spinors is
$\Delta_c=\Lambda^*(U\otimes \bC)$. This decomposes into
two complex chiral representations
according to the degree of the form $\Delta_c^+=\Lambda^{{\rm even}}(U\otimes \bC)$
and  $\Delta_c^-=\Lambda^{{\rm odd}}(U\otimes \bC)$.
These are the complex Weyl representations
of $Spin(9,1)$.
The gamma matrices are represented on $\Delta_c$ as
\bea
\Gamma_0\eta&=& -e_5\wedge\eta +e_5\lc\eta~,~~~~
\Gamma_5\eta= e_5\wedge\eta+e_5\lc \eta
\cr
\Gamma_i\eta&=& e_i\wedge \eta+ e_i\lc \eta~,~~~~~~i=1,\dots,4
\cr
\Gamma_{5+i}\eta&=& i e_i\wedge\eta-ie_i\lc\eta~.
\eea
The gamma matrices have been chosen such that
  $\{\Gamma_i; i=1,\dots, 9\}$  are Hermitian
and $\Gamma_0$ is anti-Hermitian with respect to the (auxiliary)
inner product
\be
<z^a e_a, w^b e_b>=\sum_{a=1}^{5}  (z^a)^* w^a~,~~~~
\ee
on $U\otimes \bC$  and then extended to $\Delta_c$,
where $(z^a)^*$ is the standard
complex conjugate\footnote{In (\cite{gran}) we denote
the standard complex
of $\eta$ with $\bar\eta$ instead of $\eta^*$ that we use here.} of $z^a$.
The above gamma matrices
satisfy the Clifford algebra relations
$\Gamma_A\Gamma_B+\Gamma_B \Gamma_A=2 \eta_{AB}$ with respect to the
Lorentzian inner product as expected.

The Dirac inner product on the space of spinors $\Delta_c$ is defined as
\bea
D(\eta,\theta)=<\Gamma_0\eta, \theta>~.
\eea
Observe that while $D$ is invariant under $Spin(9,1)$ the auxiliary inner product
 $<,>$ is not.

It is known that on even-dimensional manifolds, there are two Spin invariant
Majorana inner products. Only one of these inner product is Pin invariant as well.
Therefore we expect that there are two $Spin(9,1)$ invariant Majorana inner products.
One of them is defined as
\bea
A(\eta, \theta)=<A(\eta^*), \theta>~,
\eea
where the map denoted with the same symbol as the inner product is
$A=\Gamma_{12345}$. The other $Spin (9,1)$ invariant inner product is
\be
B(\eta,\theta)= <B(\eta^*), \theta>~,~~~~~~~~
\ee
where  $B=\Gamma_{06789}$. Observe that the inner product $B$ is in
addition Pin invariant and
skew-symmetric $B(\eta, \theta)=-B(\theta,\eta)$.
The above inner products $A$ and $B$  pair the $\Delta^+_c$
 and $\Delta^-_c$ representations.
Moreover, both $\Delta^+_c$ and $\Delta^-_c$ are null with respect to
 $A, B$, i.e. $A, B$
restricted to either $\Delta^+_c$ or $\Delta^-_c$ vanish.

It is well-known that $Spin(9,1)$ admits two inequivalent
Majorana-Weyl representations.
So it remains to impose the Majorana condition on the complex Weyl
representations we have
constructed above. This is done by setting  the
Dirac conjugate spinor  to be equal to the Majorana conjugate one. Equivalently,
one can impose the reality condition using an anti-linear map
 which commutes with the generators
of $Spin(9,1)$ and squares to one. There are two ways of imposing the Majorana-Weyl
condition each associated with the two Majorana inner products $A,B$ described
above. The associated anti-linear maps are
\bea
L_+&=& e^{i\varphi_+}\Gamma_0 A *
\cr
L_-&=& e^{i\varphi_-} \Gamma_0 B *~.
\la{mrc}
\eea
The phases in $L_{\pm}$ are arbitrary. Clearly $L_\pm$ are antilinear
and commute with the generators of $Spin(9,1)$. The Majorana conditions on the spinors
are
\bea
L_\pm(\eta)=\eta~.
\eea
These reality conditions map forms of even (odd)-degree to forms of even (odd)-degree
and select real subspaces $\Delta^+_{16}$ and $\Delta^-_{16}$
in   $\Delta_c^+$ and  in $\Delta^-_c$, respectively. These subspaces are the modules of
the two inequivalent  Majorana-Weyl representations of $Spin(9,1)$.

In the formulation of IIB supergravity, one can use either the $A$ ($L_+$) or $B$
($L_-$) inner
product (anti-linear reality map). In this paper, we describe our calculation
using the $B$ inner product. The reason for this is that $B$ can be extended
to the $Spin(10,1)$ invariant inner product of eleven-dimensional supergravity.
This makes connection with the description of spinor in terms of forms in
\cite{gran}.   In particular the $B$ inner product described in \cite{gran}
is equivalent up to a change of basis to the inner product $B$ that we have given above.
It is also convenient to simply somewhat the Majorana reality condition (\ref{mrc}).
In particular we choose the phase such that
\be
\eta=-\Gamma_0 B(\eta^*)~,
\ee
or equivalently
\be
\eta^*=\Gamma_{6789}\eta~.
\la{rcon}
\ee
The map $C=\Gamma_{6789}$ is also called charge conjugation matrix, $L_-=C*$.
Observe that the
anti-linear operator $C*$ commutes
with the gamma matrices, i.e. $C*\Gamma_A=\Gamma_A  C *$ or equivalently
$C^{-1} \Gamma_A C= \Gamma_A^*$. As we have seen, the reality condition
can also be expressed as
$\eta=C *\eta= C(\eta^*)$.

We shall illustrate the reality condition (\ref{rcon}) with an example.
Consider  the complex chiral spinor $a 1+ b e_{1234}$, $a,b\in \bC$.
The associated real spinor of positive chirality is
\be
\eta=a 1+  a^* e_{1234}~.
\ee
So we find  two real spinors given by $1+e_{1234}$ and $i 1-i e_{1234}$.

It remains to give the spacetime forms associated with pair of spinors $\eta,\theta$.
These are
\be
\alpha(\eta, \theta)={1\over k!} B(\eta,\Gamma_{A_1\dots A_k} \theta)
e^{A_1}\wedge\dots\wedge e^{A_k}~,~~~~~~~k=0,\dots, 9~.
\la{forms}
\ee
If both spinors are of the same chirality, then
it is sufficient to compute the forms up to degree $k\leq 5$.
This is because the forms with degrees  $k\geq 6$
are related to those with degrees $k\leq 5$ with a
Hodge duality operation. The forms of middle dimension
are either self-dual or anti-self-dual.

In many computations that follow it is convenient  to use another
 basis in the space of spinors $\Delta_c$. This
basis is given in terms of creation and annihilation
 operators. For this first write
\be
\Gamma_{\bar\a}= {1\over \sqrt {2}}(\Gamma_\a+i \Gamma_{\a+5})~,~~~~~~~~~
\Gamma_\pm={1\over \sqrt{2}} (\Gamma_5\pm\Gamma_0)
~,~~~~~~~~~\Gamma_{\a}= {1\over \sqrt {2}}(\Gamma_\a-i \Gamma_{\a+5})~.
\la{hbasis}
\ee
Observe that the Clifford algebra relations in the above basis are
$\Gamma_A\Gamma_B+\Gamma_B\Gamma_A=2 g_{AB}$,   where the non-vanishing
components of the metric are
$g_{\a\bar\b}=\delta_{\a\bar\b}, g_{+-}=1$. In addition we define
$\Gamma^B=g^{BA} \Gamma_A$.
The $1$ spinor is a Clifford  vacuum, $\Gamma_{\bar\a}1=\Gamma_+ 1=0$
and  the representation $\Delta_c$
can be constructed by acting on $1$ with the creation operators
$\Gamma^{\bar\a}, \Gamma^+$
or equivalently any spinor can be written as
\be
\eta= \sum_{k=0}^5 {1\over k!}~ \phi_{\bar a_1\dots \bar a_k}~
 \Gamma^{\bar a_1\dots\bar a_k} 1~,~~~~\bar a=\bar\a, +~,
 \la{hbasisa}
\ee
i.e. $\Gamma^{\bar a_1\dots\bar a_k} 1$, for $k=0,\dots,5$, is a basis in the
space of (Dirac) spinors.
This is another manifestation of the relation between spinors and forms. See also
\cite{oisin} for other bases of spinors used in the context of supergravity.

\subsection{Spacetime forms from spinors}

To compute the spacetime forms that are associated with the
$Spin(7)\ltimes \bR^8$- and $SU(4)\ltimes \bR^8$-invariant spinors, it is
sufficient to know the spacetime forms associated with the $1$ and $e_{1234}$ spinors. This
is because as we have seen $1$ and $e_{1234}$ span the $Spin(7)\ltimes \bR^8$- and
$SU(4)\ltimes \bR^8$-invariant spinors. As a result, the spacetime forms associated
with the $Spin(7)\ltimes \bR^8$- and $SU(4)\ltimes \bR^8$-invariant spinors
are linear combinations of the  $1$ and $e_{1234}$ spinor form bi-linears.
Using (\ref{forms}), it is easy to find that the forms associated with the
$1$ and $e_{1234}$ spinors are the following: A one-form
\bea
\kappa(e_{1234},1)=\kappa(1, e_{1234})= e^0-e^5~,
\eea
a three-form
\bea
\xi(e_{1234}, 1)=- \xi(1,e_{1234})=i (e^0-e^5)\wedge \omega~,
\eea
and  five-forms
\bea
\tau(1,1)=(e^0-e^5)\wedge \chi
\cr
\tau(e_{1234}, e_{1234})= (e^0-e^5)\wedge \chi^*
\cr
\tau(e_{1234}, 1)=\tau(1,e_{1234})=- {1\over2} (e^0-e^5)\omega\wedge \omega~,
\eea
where
\bea
\omega= e^1\wedge e^6+ e^2\wedge e^7+e^3\wedge e^8+e^4\wedge e^9
\cr
\chi=(e^1+i e^6)\wedge
(e^2+i e^7)\wedge (e^3+i e^8)\wedge (e^4+i e^9)~.
\eea
Note that $\chi$ and $\omega$ are the familiar $SU(4)$ invariant forms.

\appendix{The solution of Killing spinor equations for $SU(4)\ltimes \bR^8$-
invariant spinors}

\subsection{The solution of the linear system}

The independent components of the $P$ and $G$ fluxes  are
the following:
\be
P_+~,~~~~P_-~,~~~~P_\a~,~~~~~P_{\bar\a}
\ee
\bea
&&G_{+-\a}~,~~~~G_{+-\bar\a}~,~~~~G_{+\a\b}~,~~~~G_{+\a\bar\b}~,~~~~G_{+\bar\a\bar\b}
\cr
&&
G_{-\a\b}~,~~~~G_{-\a\bar\b}~,~~~G_{-\bar\a\bar\b}~,~~~G_{\a\b\g}~,~~~~G_{\a\b\bar\g}~,~~~
G_{\a\bar\b\bar\g}~,~~~~G_{\bar\a\bar\b\bar \g}~.
\eea
The five-form flux $F$ can be decomposed as
\bea
&&
F_{+-\a\b\g}~,~~~F_{+-\a\b\bar\g}~,~~~F_{-\a\b\g\d}~,~~~F_{-\a\b\g\bar\d}~,~~~
F_{-\a\b\bar\g\bar\d}~,
\cr
&&
F_{+\a\b\g\d}~,~~~F_{+\a\b\g\bar\d}~,~~~F_{+\a\b\bar\g\bar\d}~,~~~
F_{\a\b\g\d\bar\epsilon}~,~~~F_{\a\b\g\bar\d\bar\epsilon}~,
\eea
up to complex conjugation. However, these components are not independent but
they are related by the self-duality condition of $F$.
The self-duality condition $F_{M_1\dots M_5}=- {1\over 5!}
\epsilon_{M_1\dots M_5}{}^{N_1\dots N_5}
F_{N_1\dots N_5}$ relates the above components of $F$ as
\bea
F_{\a_1\a_2\a_3\a_4\bar\b}&=&-{1\over6} \epsilon_{\a_1\a_2\a_3\a_4}
\epsilon_{\bar\b}{}^{\g_1\g_2\g_3} F_{-+\g_1\g_2\g_3}
\cr
F_{\a_1\a_2\a_3\bar\b_1\bar\b_2}&=&-{1\over2} \epsilon_{\a_1\a_2\a_3}{}^{\bar\g_1}
\epsilon_{\bar\b_1\bar\b_2}{}^{\g_2\g_3} F_{-+\bar\g_1\g_2\g_3}
\cr
F_{+\a_1\a_2\a_3\a_4}&=&0
\cr
F_{+\a_1\a_2\a_3\bar\b}&=&-{1\over6} \epsilon_{\bar\b}{}^{\b_1\b_2\b_3}
\epsilon_{\a_1\a_2\a_3}{}^{\bar\g} F_{+\b_1\b_2\b_3\bar\g}
\cr
F_{+\bar\b_1\bar\b_2\a_1\a_2}&=&-{1\over4} \epsilon_{\bar\b_1\bar\b_2}{}^{\d_1\d_2}
\epsilon_{\a_1\a_2}{}^{\bar\g_1\bar\g_2} F_{+\bar\g_1\bar\g_2\d_1\d_2}
\cr
F_{-\a_1\a_2\a_3\a_4}&=&{1\over4!} \epsilon_{\a_1\a_2\a_3\a_4}
\epsilon^{\b_1\b_2\b_3\b_4} F_{-\b_1\b_2\b_3\b_4}
\cr
F_{-\a_1\a_2\a_3\bar\b}&=&{1\over6} \epsilon_{\bar\b}{}^{\b_1\b_2\b_3}
\epsilon_{\a_1\a_2\a_3}{}^{\bar\g} F_{-\b_1\b_2\b_3\bar\g}
\cr
F_{-\bar\b_1\bar\b_2\a_1\a_2}&=&{1\over4} \epsilon_{\bar\b_1\bar\b_2}{}^{\d_1\d_2}
\epsilon_{\a_1\a_2}{}^{\bar\g_1\bar\g_2} F_{-\bar\g_1\bar\g_2\d_1\d_2}~.
\la{selfdual}
\eea
These imply the following relations
\bea
F_{\a_1\a_2\a_3\d}{}^\d&=&F_{-+\a_1\a_2\a_3}
\cr
F_{\a_1\a_2\bar\b\d}{}^\d&=&-F_{-+\a_1\a_2\bar\b}-2 g_{\bar\b[\a_1} F_{\a_2]-+\d}{}^\d
\cr
F_{\a\b}{}^\b{}_{\g}{}^\g&=&2 F_{-+\a\d}{}^\d
\cr
F_{+\a\b\d}{}^\d&=&0
\cr
F_{+\a_1\a_2\a_3[\bar\b} \epsilon_{\bar\g]}{}^{\a_1\a_2\a_3}&=&0
\cr
F_{+\a}{}^\a{}_\b{}^\b&=&0~.
\la{trselfdual}
\eea

Since the $P$ and $G$ are complex fields, the
holomorphic and anti-holomorphic components
are not complex conjugate, i.e. $P_\a\not=({P_{\bar\a}})^*$.
Because of this, it is convenient to solve the Killing spinor equations
 for $G$ and $P$ first. The remaining equations can then be expressed
 in terms of the spacetime connection $\Omega$, the five-form flux
  $F$ and the scalar connection
 $Q$. Since $\Omega$, $F$ and $Q$ are real, one can analyze the remaining equations
 using complex conjugation. Throughout this computation, we use the
 self-duality condition of $F$.

To solve the Killing spinor equations, we assume that
 the functions $f,g_1,g_2$ are generic. In particular, we shall take
 $f\not=\pm g_2\not=0$ and $g_1\not=0$. We shall
give the solutions of some special cases in appendices C, D and E.
First consider the conditions associated with the
algebraic Killing spinor equation.
Treating (\ref{dthree}) and (\ref{dfour}) as equations for the variables
$P_+$ and $G_{+\a}{}^\a$, observe that the determinant
of coefficients is $-{1\over2} (f^2+g_1^2+g_2^2)$.
So unless the spinor $\epsilon$ vanishes, we find that
\bea
P_+= G_{+\a}{}^\a=0~.
\la{sdthreefour}
\eea
The conditions (\ref{done}) and (\ref{dtwo}) are viewed
as equations for $P_\a$ and $P_{\bar\a}$ and so both
are determined in terms of components of the $G$ flux. From now on,
we shall assume that the
$+$, $\a$ and $\bar\a$-derivatives of the scalars are
determined. $P_-$ remains undetermined by the Killing spinor
equations.

Finally, (\ref{dfive}) is viewed as an equation which relates the
$G_{+\a\b}$ and $G_{+\bar\a\bar\b}$ components
of the $G$ flux. It will be used later to  express  $G_{+\a\b}$ and
$G_{+\bar\a\bar\b}$ in terms of
components of the $F$ flux and geometry.

Dualizing (\ref{abfive}) with respect the epsilon tensor, using
(\ref{sdthreefour})  and taking a trace, we find
\bea
\Omega_{\b,+}{}^{\b}=0~.
\la{trabfive}
\eea
Substituting (\ref{trabfive}) into (\ref{pone}), we find
\bea
D_+ (f-g_2+i g_1)+ [{1\over2} \Omega_{+,\g}{}^\g+{1\over2} \Omega_{+,-+}] (f-g_2+i g_1)=0~.
\la{ponetrafive}
\eea
Substituting (\ref{trabfive}) into (\ref{pthree}) and taking the complex conjugate, we get
\bea
 D^*_+ (f+g_2-i g_1)+ [{1\over2} \Omega_{+,\g}{}^\g+{1\over2}
\Omega_{+,-+}] (f+g_2-i g_1)=0~.
\la{pthreetrabfive}
\eea
Taking the sum and the difference of the above two equations, we deduce
\bea
\partial_+ f+\frac{i}{2}(g_2-i g_1)Q_+ + [{1\over2} \Omega_{+,\g}{}^\g
+{1\over2} \Omega_{+,-+}] f=0~,
\la{pthreetrabfivea}
\eea
and
\bea
\partial_+ (g_2-i g_1)+\frac{i}{2}f Q_+ + [{1\over2} \Omega_{+,\g}{}^\g
+{1\over2} \Omega_{+,-+}] (g_2-i g_1) =0~.
\la{pthreetrabfiveb}
\eea
Taking the complex conjugate of (\ref{pthreetrabfivea}), we find that
\bea
\frac{i g_2}{f}Q_+ +\Omega_{+,\g}{}^\g=0
\eea
and
\bea
\partial_+ f+ \frac{g_1}{2}Q_+ +{1\over2} \Omega_{+,-+} f=0~.
\eea
Then (\ref{pthreetrabfiveb}) implies that
\bea
-\frac{1}{2f}(f^2-g_2^2)Q_+ +\partial_+ g_1+{1\over2} \Omega_{+,-+} g_1=0
\cr
-\frac{g_1 g_2}{2f}Q_+ +\partial_+ g_2+{1\over2} \Omega_{+,-+} g_2=0~.
\eea

Next let us turn to (\ref{afour}) and (\ref{abfive}) to determine $G_{+\a\bar\b}$.
Solving the latter in terms
of $G_{+\a\bar\b}$, we find
\bea
-{1\over2} G_{+\a\bar\b}&=&{f+g_2+ ig_1\over f-g_2- ig_1} [
\Omega_{\bar\b,+\a}- i F_{+\a\bar\b \d}{}^\d]~.
\la{afourabfive}
\eea
Next substituting $G_{+\a\bar\b}$ in (\ref{afour}),
 and taking its complex conjugate, we find that
 \bea
 \Omega_{\a,\bar\b+}+\Omega_{\bar\b,\a+}=0~,
 \la{symOp}
 \eea
 and
 \bea
 i (f^2+ g_2^2+ g_1^2) F_{+\a\bar\b\d}{}^\d&+& 2 f g_2 \Omega_{\a,+\bar\b}
 =0~.
 \la{sFpaabtr}
 \eea
This equation determines the $F_{+\a\bar\b\d}{}^\d$ component of the
flux $F$ in terms of the geometry.
Substituting  (\ref{sFpaabtr}) in (\ref{afourabfive}),
we can determine $G_{+\a\bar\b}$
in terms of the geometry as
\bea
G_{+\a\bar\b}= 2{(f+ig_1)^2- g_2^2\over f^2+g_2^2+g_1^2}\,
 \Omega_{\a,+\bar\b}~.
\la{Gpaba}
\eea

Next we turn our attention to  (\ref{afive}), (\ref{abfour}) and (\ref{dfive}).
First, we symmetrize the free indices in (\ref{abfour}) and in the dual of (\ref{afive})
 to find
\bea
F_{+\bar\b_1\bar\b_2\bar\b_3 (\a} \epsilon_{\g)}{}^{\bar\b_1\bar\b_2\bar\b_3}=0~,
\la{Fpaaasymx}
\eea
which together with the duality constraint implies
\bea
F_{+\a_1\a_2\a_3\bar\b}=0~,
\la{Fpaaasym}
\eea
and
\bea
\Omega_{\a,+\b}+\Omega_{\b,+\a}=0~.
\la{Opaa}
\eea
Next, we dualize (\ref{afive}) and take the difference with the dual of (\ref{abfour})
in such a way as to eliminate the $G$ dependence  and to find
\bea
 (f+g_2+i g_1) \Omega_{\a,+\b}- {1\over2} (f-g_2+ ig_1)
\epsilon_{\a\b}{}^{\bar \g_1\bar\g_2} \Omega_{\bar\g_1,+\bar\g_2}=0~.
\la{abfourdafive}
\eea
Comparing this expression with its complex conjugate, we get that
\bea
\Omega_{\a,+\b}=0~.
\la{Fpaatr}
\eea

Taking the sum and the difference of (\ref{dfive}) and (\ref{abfour})
and comparing the two,  we deduce that
\bea
 G_{+\bar\a\bar\b}=0~,
\la{Gpabab}
\eea
and
\bea
 G_{\a\b+}=0~.
 \la{Gpaa}
 \eea

To continue, we substitute (\ref{Gpabab}) and (\ref{Gpaa}) into (\ref{ptwo})
and then use (\ref{Fpaatr}). After some computation, we find
that
\bea
 \Omega_{+,\bar\a\bar\b}=0~.
 \eea
This relates the $\Omega_{+,\bar\a\bar\b}$ component of the connection
to the $\Omega_{\g_1,+\g_2}$ component.

Next, we multiply
(\ref{mone}) with $f-g_2-ig_1$ and (\ref{mthree}) with $f+g_2-i g_1$ and take
their sum. Separating the resulting expression in real and imaginary parts, we  get
\bea
\partial_- (f^2+g_2^2+ g_1^2)+ \Omega_{-,-+}(f^2+g_2^2+ g_1^2)=0~,
\eea
and
\bea
-i(f^2+g_1^2+g_2^2) Q_-+ 2i f\partial_-g_1- 2i g_1\partial_-f
-2f g_2 \Omega_{-,\a}{}^\a+ {i\over2} F_{-\a}{}^\a{}_\b{}^\b (f^2+g_2^2+ g_1^2)
\cr
+{i\over12} [f^2-(g_2+ig_1)^2] F_{-\g_1\g_2\g_3\g_4} \epsilon^{\g_1\g_2\g_3\g_4}
\cr
+{i\over12} [f^2-(g_2-ig_1)^2] F_{-\bar\g_1\bar\g_2\bar\g_3\bar\g_4}
 \epsilon^{\bar\g_1\bar\g_2\bar\g_3\bar\g_4}=0~.
 \la{Fmdcontr}
 \eea
This equation can be thought of as determining $ F_{-\a}{}^\a{}_\b{}^\b$ in terms
of the other fluxes and geometry. The component $F_{-\bar\g_1\bar\g_2\bar\g_3\bar\g_4}$
and therefore also its conjugate is
not specified by the Killing spinor equations.
We multiply
(\ref{mone}) with $f+g_2+ig_1$ and (\ref{mthree}) with $f-g_2+i g_1$  and take
the difference. This gives
\bea
-2f \partial_-g_2+ 2 g_2\partial_-f+ 2i g_2\partial_-g_1-2i g_1\partial_-g_2
+\Omega_{-,\a}{}^\a [(f+i g_1)^2- g_2^2]
\cr
+ {1\over2} (f^2+g_2^2+ g_1^2)
G_{-\a}{}^\a+ {i\over12} (f+g_2+i g_1)^2 F_{-\g_1\g_2\g_3\g_4}
\epsilon^{\g_1\g_2\g_3\g_4}
\cr
-{i\over12}
(f-g_2+i g_1)^2 F_{-\bar\g_1\bar\g_2\bar\g_3\bar\g_4}
\epsilon^{\bar\g_1\bar\g_2\bar\g_3\bar\g_4}=0~.
\la{Gmcontr}
\eea
This equation expresses the component  $G_{-\a}{}^\a$ of the $G$ flux
in terms of the $F$ flux and geometry.

The equation (\ref{mtwo}) contains the components $F_{-\a\b\d}{}^\d$,
$G_{-\a\b}$ and $G_{-\bar\a\bar\b}$. So it can be used to determine
either $G_{-\a\b}$ or $G_{-\bar\a\bar\b}$ in terms of the other two components.
There is no obvious advantage to give explicitly the
 solution since the remaining fluxes are not
determined by the rest of the equations of the linear system.

The conditions (\ref{atwo}), (\ref{abone}),
(\ref{abthree}) and (\ref{mfour}) should be investigated together.
First, we take the trace of (\ref{atwo}) to find
\bea
(f-g_2+ig_1)
[\Omega_{\b,}{}^\b{}_{\bar\a}-i
F_{-+\b}{}^\b{}_{\bar\a}]  + (f+g_2-i g_1) [{1\over8}G_{\bar\a\b}{}^\b-{3\over8}
G_{\bar\a-+}]
\cr
-
 {1\over2}(f+g_2+ig_1)
\Omega_{\g_1,\g_2\g_3} \epsilon^{\g_1\g_2\g_3}{}_{\bar\a}
-{1\over8} (f-g_2-i g_1) G_{\g_1\g_2\g_3} \epsilon^{\g_1\g_2\g_3}{}_{\bar\a}=0~.
\eea
Combining the above equation with (\ref{mfour}), we get
\bea
(f-g_2+ig_1) [\Omega_{\b,}{}^\b{}_{\bar\a}+
\Omega_{-,+\bar\a}]
-(f+g_2+ ig_1) [{1\over2} \Omega_{\g_1,\g_2\g_3}
+{i\over3} F_{\g_1\g_2\g_3-+}] \epsilon^{\g_1\g_2\g_3}{}_{\bar\a}
\cr
-{1\over12} (f-g_2-i g_1) G_{\g_1\g_2\g_3}\epsilon^{\g_1\g_2\g_3}{}_{\bar\a}=0~,
\la{Gaaa}
\eea
which determines $G_{\g_1\g_2\g_3}$ in terms of the $F$ flux and geometry.
Substituting the above solution for $G_{\g_1\g_2\g_3}$ in both
(\ref{mfour}) and (\ref{abone}), and summing them with appropriate numerical
factors, we get
\bea
D_{\bar\a} (f-g_2+i g_1)+ [{1\over2} \Omega_{\bar\a,\b}{}^\b+ {1\over2}
\Omega_{\bar\a,-+}+\Omega_{\b,}{}^\b{}_{\bar\a}
+ i F_{-+\bar\a\b}{}^\b
+2\Omega_{-,+\bar\a}] (f-g_2+ i g_1)
\cr
+(f+g_2+ ig_1)[-{1\over2}\Omega_{\g_1,\g_2\g_3}-{i\over3}
F_{\g_1\g_2\g_3-+}]\epsilon^{\g_1\g_2\g_3}{}_{\bar\a}+ {1\over2} G_{-+\bar\a} (f+g_2- ig_1)=0~.
\la{Gpmab}
\eea
This  equation  determines $G_{-+\bar\a}$ in terms
of the $F$ fluxes and the geometry.
If instead, we take the difference of (\ref{abone}) and  (\ref{Gpmab})
with appropriate numerical factors,
we get
\bea
{3\over4} D_{\bar\a} (f-g_2+ i g_1)+[ {3\over8} \Omega_{\bar\a,\b}{}^\b+
{3\over8} \Omega_{\bar\a,-+}
\cr
-{i\over4} F_{-+\bar\a\b}{}^\b+{1\over4} \Omega_{\b,}{}^\b{}_{\bar\a}]
(f-g_2+i g_1)+ {1\over8} G_{\bar\a\b}{}^\b
(f+ g_2- ig_1)
\cr
+ (f+ g_2+ ig_1) [-{1\over8} \Omega_{\g_1,\g_2\g_3} +{i\over4} F_{\g_1\g_2\g_3-+}]
\epsilon^{\g_1\g_2\g_3}{}_{\bar\a}=0~,
\la{Gatr}
\eea
which determines the component $G_{\bar\a\b}{}^\b$ of $G$. By
substituting the above results into (\ref{atwo}), we can  solve
for $G_{\a\bar\b\bar\g}$
\bea
G_{\a\bar\b\bar\g}= \frac{1}{(f+g_2-i
g_1)}\Big(-4g_{\a[\bar\b}D_{\bar\g]}(f-g_2+i g_1)+(f-g_2+i
g_1)(-2\Om_{\a,\bar\b\bar\g}
\cr
-2g_{\a[\bar\b}\Om_{\bar\g],\d}{}^\d+2g_{\a[\bar\b}\Om_{\bar\g],+-}-4i
F_{-+\a\bar\b\bar\g}+4ig_{\a[\bar\b}F_{\bar\g]-+\d}{}^\d)
\cr
+(f+g_2+i
g_1)\ep_{\bar\b\bar\g}{}^{\d_1\d_2}(\Om_{\a,\d_1\d_2}-2i F_{-+\a\d_1\d_2})
\Big)~.
\la{Gaabab}
\eea
We have thus solved all three equations (\ref{atwo}), (\ref{abone}),
 and (\ref{mfour}) for $G_{\a\b\g}$, $G_{-+\bar\a}$, $G_{\bar\a\b}{}^\b$
 and $G_{\a\bar\b\bar\g}$.

 It remains to solve (\ref{abthree}). For this first,  we compute the difference
 (\ref{Gatr}) and  (\ref{Gpmab}) with appropriate numerical factors and find
 \bea
 D_{\bar\a}(f-g_2+ ig_1)+[ {1\over2} \Omega_{\bar\a,\b}{}^\b+ {1\over2}
 \Omega_{\bar\a,-+}
 -i
 F_{-+\bar\a\b}{}^\b
 \cr
 - \Omega_{-,+\bar\a}] (f-g_2+i g_1)
 +[-{1\over4} G_{-+\bar\a}+{1\over4} G_{\bar\a\b}{}^\b] (f+g_2-i g_1)
 \cr
 +{2i\over3}(f+g_2+i g_1)
  F_{\g_1\g_2\g_3-+} \epsilon^{\g_1\g_2\g_3}{}_{\bar\a}=0~.
 \la{GatrGpmab}
 \eea
 Next we multiply (\ref{GatrGpmab}) with $f-g_2-ig_1$ and (\ref{abthree})
 with $f+g_2-ig_1$ to  get
 \bea
 \partial_{\bar\a} (f^2+ g_2^2+ g_1^2)
 -i Q_{\bar\a}(f^2+ g_2^2+ g_1^2)+ 2i f D_{\bar\a} g_1- 2i g_1D_{\bar\a} f
 +[\Omega_{\bar\a, -+}
 \cr
  - 2i F_{-+\bar\a\b}{}^\b
 - \Omega_{-,+\bar\a}] (f^2+ g_2^2+ g_1^2)
 -2 f g_2 [\Omega_{\bar\a,\b}{}^\b
 -\Omega_{-,+\bar\a}]
 \cr
 +  {2i\over3} (f^2-(g_2+ig_1)^2)
 F_{\g_1\g_2\g_3-+}
 \epsilon^{\g_1\g_2\g_3}{}_{\bar\a}=0~.
 \la{Fabaaaax}
 \eea
 This will be compared later with another
 equation which we shall derive by examining the rest of the equations.

 Similarly, the conditions (\ref{aone}), (\ref{athree}), (\ref{abtwo})
 and (\ref{mfive}) should be investigated together.  We first dualize (\ref{abtwo})
 and then take the trace to find
 \bea
 (f+g_2+ig_1) [-\Omega_{\bar\b,}{}^{\bar\b}{}_\a
 -i F_{-+\a\b}{}^\b]
 + (f-g_2-i g_1) [{1\over8} G_{\a\b}{}^\b+{3\over8} G_{\a-+}]
 \cr
 +{1\over2}(f-g_2+ig_1)  \Omega_{\bar\g_1,\bar\g_2\bar\g_3}
 \epsilon^{\bar\g_1\bar\g_2\bar\g_3}{}_\a
 +\frac{1}{8}(f+g_2-ig_1)G_{\bar\g_1\bar\g_2\bar\g_3}\ep^{\bar\g_1\bar\g_2\bar\g_3
}{}_{\a} =0~.
 \la{abtwotr}
\eea
We take the sum of (\ref{abtwotr}) with  the dual of (\ref{mfive}) after weighting
the equations with appropriate coefficients to get
\bea
(f+g_2+i g_1) [-\Omega_{\bar\b,}{}^{\bar \b}{}_\a-\Omega_{-,+\a}]
+(f-g_2+i g_1)[{1\over2} \Omega_{\bar\g_1, \bar\g_2\bar\g_3}
+{i\over3} F_{-+\bar\g_1 \bar\g_2\bar\g_3}]
\epsilon^{\bar\g_1 \bar\g_2\bar\g_3}{}_{\a}
\cr
+{1\over12} (f+g_2-ig_1) G_{\bar\g_1 \bar\g_2\bar\g_3}
\epsilon^{\bar\g_1 \bar\g_2\bar\g_3}{}_{\a}
=0~.
\la{Gababab}
\eea
This equation gives $G_{\bar\g_1 \bar\g_2\bar\g_3}$
in terms of the $F$ flux and the geometry.

We substitute $ G_{\bar\g_1 \bar\g_2\bar\g_3}$ in
 the dual of (\ref{mfive}) and in (\ref{athree})
and take their difference after weighting the equations with appropriate coefficients
 to find
\bea
D_\a (f+g_2+i g_1)+ [-{1\over2} \Omega_{\a,\b}{}^\b+{1\over2} \Omega_{\a,-+}
-i F_{-+\a\g}{}^\g
\cr
+\Omega_{\bar\b,}{}^{\bar\b}{}_\a+2\Omega_{-,+\a}]
(f+ g_2+i g_1)
+{1\over2} G_{-+\a} (f-g_2-ig_1)
\cr
+(f-g_2+i g_1)[-{1\over2}
\Omega_{\bar\g_1,\bar\g_2\bar\g_3}-{i\over3}
 F_{-+\bar\g_1\bar\g_2\bar\g_3}]\epsilon^{\bar\g_1\bar\g_2\bar\g_3}{}_\a=0~,
\la{Gmpab}
\eea
which gives $G_{-+\a}$ in terms of the other
fluxes and geometry. By substituting the above results into
(\ref{abtwo}), we can now solve for $G_{\a\b\bar\g}$ to get
\bea
G_{\a\b\bar\g}=\frac{1}{(f-g_2-i
g_1)}\Big(-4g_{\bar\g[\a}D_{\b]}(f+g_2+i g_1)+(f+g_2+i
g_1)(-2\Om_{\bar\g,\a\b}
\cr
+2g_{\bar\g[\a}\Om_{\b],\d}{}^\d+2g_{\bar\g[\a}\Om_{\b],+-}-4i
F_{-+\a\b\bar\g}-4ig_{\bar\g[\a}F_{\b]-+\d}{}^\d)
\cr
+(f-g_2+i
g_1)\ep_{\a\b}{}^{\bar\d_1\bar\d_2}(\Om_{\bar\g,\bar\d_1\bar\d_2}-2i F_{-+\bar\g\bar\d_1\bar\d_2})
\Big)~.
\la{Gaaab}
\eea

Subtracting (\ref{Gmpab}) from (\ref{athree}) with appropriate factors,
we find
\bea
{3\over4} D_\a(f+g_2+ig_1) +[-{3\over8} \Omega_{\a,\b}{}^\b+{3\over8}
\Omega_{\a,-+}+{i\over4}
F_{-+\a\d}{}^\d
\cr
+{1\over4} \Omega_{\bar\b,}{}^{\bar\b}{}_\a] (f+g_2+i g_1)
-{1\over8} G_{\a\d}{}^\d (f-g_2-ig_1)
\cr
+ (f-g_2+ig_1) [-{1\over8} \Omega_{\bar\g_1,\bar\g_2\bar\g_3} +{i\over4}
F_{-+\bar\g_1\bar\g_2\bar\g_3}]
\epsilon^{\bar\g_1\bar\g_2\bar\g_3}{}_\a=0~,
\la{Gabtr}
\eea
which gives $G_{\a\d}{}^\d$ in terms of the $F$ flux and geometry.

Subtracting (\ref{Gmpab}) from  (\ref{Gabtr}) with appropriate numerical factors,
we find
\bea
D_\a(f+g_2+i g_1)+[-{1\over2} \Omega_{\a,\b}{}^\b+{1\over2} \Omega_{\a,-+}
+ i F_{-+\a\d}{}^\d
- \Omega_{-,+\a}] (f+g_2+i g_1)
\cr
+ [-{1\over4} G_{\a\g}{}^\g-{1\over4} G_{-+\a}] (f-g_2-ig_1)
+{2i\over3}(f-g_2+i g_1)
F_{-+\bar\g_1\bar\g_2\bar\g_3} \epsilon^{\bar\g_1\bar\g_2\bar\g_3}{}_\a=0~.
\la{GabtrGmpab}
\eea
We multiply (\ref{aone}) with $f-g_2-ig_1$ and (\ref{GabtrGmpab}) with $f+g_2-ig_1$
and sum them together. Then we take the complex conjugate
of the resulting expression to find
\bea
\partial_{\bar\a} (f^2+g_2^2+g_1^2)+i Q_{\bar\a}(f^2+g_2^2+g_1^2)-
2i f {\bar D}_{\bar\a} g_1+ 2i
g_1 {\bar D}_{\bar\a}f
\cr
+[\Omega_{\bar\a,-+}+2 i F_{-+\bar\a\b}{}^\b
- \Omega_{-,+\bar\a}] (f^2+g_2^2+g_1^2)
-2 fg_2 [-\Omega_{\bar\a,\b}{}^\b+  \Omega_{-,+\bar\a}]
\cr
- {2i\over3}(f^2- (g_2+i g_1)^2)
 F_{-+\g_1\g_2\g_3} \epsilon^{\g_1\g_2\g_3}{}_{\bar\a}=0~.
\la{Fabaaaay}
\eea
It remains to compare (\ref{Fabaaaax}) with (\ref{Fabaaaay}).  Taking the
sum and the difference, we find
\bea
\partial_{\bar\a} (f^2+g_2^2+g_1^2)-[\Omega_{\bar\a,+-}
+\Omega_{-,+\bar\a}] (f^2+g_2^2+g_1^2)=0~,
\eea
and
\bea
-2i (f^2+g_2^2+ g_1^2)Q_{\bar\a}+4i f\partial_{\bar\a} g_1-4i
g_1\partial_{\bar\a} f-4i
F_{-+\bar\a\b}{}^\b (f^2+g_2^2+g_1^2)
\cr
-4f g_2[\Omega_{\bar\a,\b}{}^\b- \Omega_{-,+\bar\a}] +{4i\over3}(
f^2- (g_2+ ig_1)^2) F_{-+\g_1\g_2\g_3}
\epsilon^{\g_1\g_2\g_3}{}_{\bar\a}=0~.
\la{Fpmaaa}
\eea
These last equation can be used to determine one more component of the $F$ flux,
say $F_{-+\g_1\g_2\g_3}$. We shall not substitute $F_{-+\g_1\g_2\g_3}$ back into the
equations that determine the $G$ fluxes. This is because the resulting equations
do not exhibit any apparent simplification. So, we shall take the scalar
fluxes to depend on the $G$ and so implicitly on the  $F$ fluxes and geometry,
the $G$ fluxes to depend on the $F$ fluxes and geometry,
 and the $F$ fluxes to depend on the geometry. The equations that
 determine the various
 components of the fluxes are summarized in the tables below.

The scalar fluxes $P$ are given in the following equations
\begin{equation}
\begin{array}{|c|c|}\hline
\mathrm{Fluxes } & \mathrm{Equations}
 \\
 \hline
P_+&(\ref{sdthreefour})  \\
P_\a   &  (\ref{dtwo})\\
P_{\bar\a}& (\ref{done})
\\ \hline
\end{array}
\end{equation}
The $G$ fluxes are determined by the following equations
\begin{equation}
\begin{array}{|c|c|}\hline
\mathrm{Fluxes } & \mathrm{Equations}
 \\\hline
G_{+\a}{}^\a&(\ref{sdthreefour})  \\
G_{+\a\bar\b}   & (\ref{Gpaba}) \\
G_{+\a\b}& (\ref{Gpaa}) \\
G_{+\bar\a\bar\b}& (\ref{Gpabab}) \\
G_{-\a}{}^\a& (\ref{Gmcontr})\\
G_{-\a\b}& (\ref{mtwo}) \\
G_{\a\b\g}&  (\ref{Gaaa}) \\
G_{-+\bar\a}& (\ref{Gpmab}) \\
G_{\a\bar\b\bar\g}& (\ref{Gaabab}) \\
G_{\bar\a\bar\b\bar\g}& (\ref{Gababab}) \\
G_{-+\a}& (\ref{Gmpab}) \\
G_{\a\b\bar\g}& (\ref{Gaaab})
\\ \hline
\end{array}
\end{equation}

The $F$ fluxes are determined by the following equations
\begin{equation}
\begin{array}{|c|c|}\hline
\mathrm{Fluxes } & \mathrm{Equations}
 \\\hline
F_{+\a\b\g\d}& (\ref{selfdual})  \\
F_{+\a\bar\b\g}{}^\g  & (\ref{sFpaabtr}) \\
F_{+\a\b\g\bar\d}& (\ref{Fpaaasym}) \\
F_{-\a}{}^\a{}_\b{}^\b& (\ref{Fmdcontr}) \\
F_{-+\a\b\g}& (\ref{Fpmaaa})
\\ \hline
\end{array}
\end{equation}
The fluxes that are not mentioned in the above tables are not restricted by the
Killing spinor equations.  $F$ is further restricted by the self-duality condition.
The conditions on the geometry have been summarized in section three.

\appendix{The solution of Killing spinor equations for the $Spin(7)\ltimes \bR^8$-invariant
 spinor}

The analysis of the conditions of the Killing spinor equations for a
$Spin(7)\ltimes \bR^8$ invariant spinor is similar to that of an
$SU(4)\ltimes \bR^8$ invariant spinor but there are some
differences. Because of this and for stating the conditions for the
existence of a parallel $Spin(7)\ltimes \bR^8$ spinor, we shall
repeat the analysis from the beginning.

The algebraic Killing spinor equations (\ref{dthree}) and (\ref{dfour}) imply
\bea
P_+=G_{+\a}{}^\a=0~,
\la{firstcond}
\eea
and (\ref{dfive}) gives
\bea
G_{+\bar\a\bar\b}-{1\over2} \epsilon_{\bar\a\bar\b}{}^{\g\d} G_{+\g\d}=0~.
\eea
The two other equations can be thought of as determining the scalars in terms of
the $G$ fluxes.

Next consider (\ref{abfive}) and (\ref{afour}). After taking the dual of the former
and the trace, and using the self-duality of $F$, we find
\bea
\Omega_{\a,+}{}^\a=0~.
\eea

Using the above equations, we take the sum and the difference of
(\ref{afour}) with the dual of (\ref{abfive}), and separate the sum in real
and imaginary parts to find
\bea
\Omega_{\a,+\bar\b}+\Omega_{\bar\b,+\a}=0~,
\eea
and
\bea
 F_{+\a\bar\b \d}{}^\d=0~.
 \la{sFpaacontr}
 \eea
 The difference, instead, gives
 \bea
 (f-ig) G_{+\a\bar\b}- 2(f+ig) \Omega_{\a,+\bar\b}=0~.
 \la{spGpaab}
 \eea
Using (\ref{pone}) and (\ref{pthree}) and
 some of the above equations,  we
 find that
 \bea
 \Omega_{+,\g}{}^\g=0~,
 \eea
 and
 \bea
g Q_+ + 2 \partial_+ f+\Omega_{+,-+} f=0~,
 \cr
-f Q_+ + 2 \partial_+ g+\Omega_{+,-+} g=0~.
\eea

We now consider (\ref{dfive}), (\ref{afive}) and (\ref{abfour}). Symmetrizing
(\ref{afive}) and (\ref{abfour}) and comparing them, we get
\bea
\Omega_{(\a,\b)+}=0~,~~~~~
F_{+\bar\g_1\bar\g_2\bar\g_3(\a} \epsilon_{\b)}{}^{\bar\g_1\bar\g_2\bar\g_3}=0~.
\la{symcondx}
\eea
The latter condition together with the self-duality of $F$ imply
\bea
F_{\a_1\a_2\a_3\bar\b}=0~.
\la{symcond}
\eea
Therefore the symmetric part of (\ref{abfour}) and of the dual of (\ref{afive})
vanishes identically. Next, we take the difference of (\ref{afive}) with
(\ref{abfour}) with appropriate factors to find
\bea
  \Omega_{\a,+\b}-{1\over2}
 \epsilon_{\a\b}{}^{\bar\g_1\bar\g_2} \Omega_{\bar\g_1,+\bar\g_2}=0~.
\eea

The conditions (\ref{dfive}), (\ref{afive}) and (\ref{abfour}) give
\bea
G_{+\a\b}-2{f+ig\over f-ig} \Omega_{\a,+\b}=0~,
\la{sGpaa}
\eea
and
\bea
G_{+\bar\a\bar\b}-2{f+ig\over f-ig} \Omega_{\bar\a,+\bar\b}=0~,
\la{sGpabab}
\eea
which determine the components of the $G$ flux in terms of the geometry.

Next let us turn our attention to (\ref{ptwo}). Using (\ref{sGpaa}) and
(\ref{sGpabab}), we find that
\bea
\Omega_{+,\bar\a\bar\b}-{1\over2} \epsilon_{\bar\a\bar\b}{}^{\g_1\g_2}
 \Omega_{+,\bar\g_1\bar\g_2}=0~.
 \eea
Taking
the difference of (\ref{mone}) and (\ref{mthree}), we get
\bea
{1\over2} G_{-\g}{}^\g (f-ig)+ \Omega_{-\g}{}^\g (f+ig)
+{i\over12}(f+i g) [ F_{-\g_1\g_2\g_3\g_4}
\epsilon^{\g_1\g_2\g_3\g_4}
\cr-
F_{-\bar\g_1\bar\g_2\bar\g_3\bar\g_4}
\epsilon^{\bar\g_1\bar\g_2\bar\g_3\bar\g_4}]=0~,
\la{sGmtr}
\eea
which can be used to determine $G_{-\g}{}^\g$. Then  splitting the sum
of (\ref{mone}) and (\ref{mthree})
in real and imaginary pieces, we find
\bea
2\partial_-f+ Q_-g+\Omega_{-,-+} f -{1\over2} F_{-\g}{}^\g{}_\d{}^\d g
-{1\over12} g [ F_{-\g_1\g_2\g_3\g_4} \epsilon^{\g_1\g_2\g_3\g_4}
\cr
+ F_{-\bar\g_1\bar\g_2\bar\g_3\bar\g_4} \epsilon^{\bar\g_1\bar\g_2\bar\g_3\bar\g_4}]=0~,
\eea
and
\bea
2\partial_- g- Q_- f+\Omega_{-,-+} g+{f\over2} F_{-\g}{}^\g{}_\d{}^\d
+{f\over12} [ F_{-\g_1\g_2\g_3\g_4} \epsilon^{\g_1\g_2\g_3\g_4}
\cr
+ F_{-\bar\g_1\bar\g_2\bar\g_3\bar\g_4} \epsilon^{\bar\g_1\bar\g_2\bar\g_3\bar\g_4}]=0~.
\eea
The last two equations give
\bea
\partial_-(f^2+g^2)+\Omega_{-,-+}(f^2+g^2)=0~,
\eea
and
\bea
2g \partial_-f-2f \partial_-g+ Q_- (f^2+g^2)-{1\over2} F_{-\g}{}^\g{}_\d{}^\d
(g^2+f^2)
\cr
-{f^2+g^2\over12} [ F_{-\g_1\g_2\g_3\g_4} \epsilon^{\g_1\g_2\g_3\g_4}
+ F_{-\bar\g_1\bar\g_2\bar\g_3\bar\g_4} \epsilon^{\bar\g_1\bar\g_2\bar\g_3\bar\g_4}]=0~.
\la{sFmtrtr}
\eea
The latter can be used to determine $F_{-\g}{}^\g{}_\d{}^\d$.

The condition (\ref{mtwo}) can be written as
\bea
[\Omega_{-,\bar\b_1\bar\b_2}-{1\over2}
\epsilon_{\bar\b_1\bar\b_2}{}^{\g_1\g_2} \Omega_{-,\g_1\g_2}+
i(F_{-\bar\b_1\bar\b_2\d}{}^\d +{1\over2}
\epsilon_{\bar\b_1\bar\b_2}{}^{\g_1\g_2} F_{-\g_1\g_2\d}{}^\d)]
(f+ig)
\cr
+{(f-ig)\over2}( G_{-\bar\b_1\bar\b_2}-{1\over2}
\epsilon_{\bar\b_1\bar\b_2}{}^{\g_1\g_2} G_{-\g_1\g_2})=0~
\la{sGmabab}
\eea
and can be used to determine the complex anti-self dual part of $G_{-\bar\a\bar\b}$.

The equations (\ref{atwo}), (\ref{abone}), (\ref{abthree}) and
(\ref{mfour}) can be analyzed as in the case with an $SU(4)\ltimes
\bR^8$ invariant spinor. We shall not give the details but instead
we shall state the results. In particular $G_{\a\b\g}$ is determined
by the equation
\bea
(f+ig) [\Omega_{\b,}{}^\b{}_{\bar\a}+ \Omega_{-,+\bar\a}]  -(f+ ig)
[{1\over2} \Omega_{\g_1,\g_2\g_3}
+{i\over3} F_{\g_1\g_2\g_3-+}] \epsilon^{\g_1\g_2\g_3}{}_{\bar\a}
\cr -{1\over12} (f-i g)
G_{\g_1\g_2\g_3}\epsilon^{\g_1\g_2\g_3}{}_{\bar\a}=0~,
\la{sGaaax}
\eea
$G_{-+\bar\a}$ is determined by
\bea
D_{\bar\a} (f+i g)+ [{1\over2} \Omega_{\bar\a,\b}{}^\b+ {1\over2}
\Omega_{\bar\a,-+}+\Omega_{\b,}{}^\b{}_{\bar\a}+ i F_{-+\bar\a\b}{}^\b
+2\Omega_{-,+\bar\a}] (f+ i g)
\cr
+(f+ ig)[-{1\over2}\Omega_{\g_1,\g_2\g_3}-{i\over3}
F_{\g_1\g_2\g_3-+}]\epsilon^{\g_1\g_2\g_3}{}_{\bar\a}+ {1\over2} G_{-+\bar\a} (f- ig)=0~,
\la{sGmpabx}
\eea
and $G_{\bar\a\b}{}^\b$ is determined by
\bea
{3\over4} D_{\bar\a} (f+ i g)+[ {3\over8} \Omega_{\bar\a,\b}{}^\b+
{3\over8} \Omega_{\bar\a,-+}
-{i\over4} F_{-+\bar\a\b}{}^\b+{1\over4}
\Omega_{\b,}{}^\b{}_{\bar\a}] (f+i g)
\cr
+ (f+ ig) [-{1\over8} \Omega_{\g_1,\g_2\g_3} +{i\over4} F_{\g_1\g_2\g_3-+}]
\epsilon^{\g_1\g_2\g_3}{}_{\bar\a}+ {1\over8} G_{\bar\a\b}{}^\b
(f- ig)=0~.
\la{sGabtrx}
\eea
Substituting the above results into (\ref{atwo}), we can  solve
for $G_{\a\bar\b\bar\g}$ to get
\bea
G_{\a\bar\b\bar\g}= \frac{1}{(f-i
g)}\Big(-4g_{\a[\bar\b}D_{\bar\g]}(f+i g)+(f+i
g)(-2\Om_{\a,\bar\b\bar\g}
\cr
-2g_{\a[\bar\b}\Om_{\bar\g],\d}{}^\d+2g_{\a[\bar\b}\Om_{\bar\g],+-}-4i
F_{-+\a\bar\b\bar\g}+4ig_{\a[\bar\b}F_{\bar\g]-+\d}{}^\d)
\cr
+(f+i
g)\ep_{\bar\b\bar\g}{}^{\d_1\d_2}(\Om_{\a,\d_1\d_2}-2i F_{-+\a\d_1\d_2})
\Big)~.
\la{sGaabab}
\eea

Similarly for the equations (\ref{aone}), (\ref{athree}), (\ref{abtwo})
and (\ref{mfive}), we find that $G_{\bar\a\bar\b\bar\g}$ is determined by
\bea
(f+i g) [-\Omega_{\bar\b,}{}^{\bar \b}{}_\a-\Omega_{-,+\a}]
+(f+i g)[{1\over2} \Omega_{\bar\g_1, \bar\g_2\bar\g_3} +{i\over3} F_{-+\bar\g_1
\bar\g_2\bar\g_3}] \epsilon^{\bar\g_1 \bar\g_2\bar\g_3}{}_{\a}
\cr
+{1\over12} (f-ig) G_{\bar\g_1 \bar\g_2\bar\g_3}\epsilon^{\bar\g_1
\bar\g_2\bar\g_3}{}_{\a} =0~,
  \la{sGabababx}
\eea
$G_{-+\a}$ is determined by
\bea
D_\a (f+i g)+ [-{1\over2} \Omega_{\a,\b}{}^\b+{1\over2}
\Omega_{\a,-+}  -i
F_{-+\a\g}{}^\g
\cr
+\Omega_{\bar\b,}{}^{\bar\b}{}_\a+2\Omega_{-,+\a}] (f+i g)
+{1\over2} G_{-+\a} (f-ig)
\cr
+(f+i g)[-{1\over2} \Omega_{\bar\g_1,\bar\g_2\bar\g_3}-{i\over3}
 F_{-+\bar\g_1\bar\g_2\bar\g_3}]\epsilon^{\bar\g_1\bar\g_2\bar\g_3}{}_\a=0~,
 \la{sGpma}
\eea
and $G_{\a\d}{}^\d$ is determined by
\bea
{3\over4} D_\a(f+ig) +[-{3\over8} \Omega_{\a,\b}{}^\b+{3\over8}
\Omega_{\a,-+}+{i\over4}
F_{-+\a\d}{}^\d
+{1\over4} \Omega_{\bar\b,}{}^{\bar\b}{}_\a] (f+i g)
\cr
+ (f+ig) [-{1\over8} \Omega_{\bar\g_1,\bar\g_2\bar\g_3}  +{i\over4}
F_{-+\bar\g_1\bar\g_2\bar\g_3}]
\epsilon^{\bar\g_1\bar\g_2\bar\g_3}{}_\a-{1\over8}
G_{\a\d}{}^\d (f-ig)=0~.
\eea
By substituting the above results into (\ref{abtwo}), we can
solve for $G_{\a\b\bar\g}$ to get
\bea
G_{\a\b\bar\g}=\frac{1}{(f-i g)}\Big(-4g_{\bar\g[\a}D_{\b]}(f+i
g)+(f+i g)(-2\Om_{\bar\g,\a\b}
\cr
+2g_{\bar\g[\a}\Om_{\b],\d}{}^\d+2g_{\bar\g[\a}\Om_{\b],+-}-4i
F_{-+\a\b\bar\g}-4ig_{\bar\g[\a}F_{\b]-+\d}{}^\d)
\cr
+(f+i
g)\ep_{\a\b}{}^{\bar\d_1\bar\d_2}(\Om_{\bar\g,\bar\d_1\bar\d_2}-
2i F_{-+\bar\g\bar\d_1\bar\d_2})
\Big)~.
\la{sGaaab}
\eea

Using the above results, (\ref{aone}) and (\ref{abthree})
yield
\bea
\partial_{\bar\a} (f^2+g^2)-[\Omega_{\bar\a,+-}+\Omega_{-,+\bar\a}] (f^2+g^2)
=0~,
\eea
and
\bea
4i f\partial_{\bar\a} g-4i g\partial_{\bar\a} f+ [-2i Q_{\bar\a}-4i F_{-+\bar\a\b}{}^\b] (f^2+g^2)
\cr
 +{4i\over3}(
f^2+g^2) F_{-+\g_1\g_2\g_3}
\epsilon^{\g_1\g_2\g_3}{}_{\bar\a}=0~.
\la{Fpmaaa2}
\eea
The last equation can be used to determine  $F_{-+\g_1\g_2\g_3}$.
We summarize below the equations that  give the components of the fluxes
as in the $SU(4)\ltimes \bR^8$ case.

The scalar fluxes $P$ are given in the following equations
\begin{equation}
\begin{array}{|c|c|}\hline
\mathrm{Fluxes } & \mathrm{Equations}
 \\
 \hline
P_+&(\ref{firstcond})  \\
P_\a   &  (\ref{dtwo})\\
P_{\bar\a}& (\ref{done})
\\ \hline
\end{array}
\end{equation}
The $G$ fluxes are determined by the following equations
\begin{equation}
\begin{array}{|c|c|}\hline
\mathrm{Fluxes } & \mathrm{Equations}
 \\\hline
G_{+\a}{}^\a&(\ref{firstcond})  \\
G_{+\a\bar\b}   &  (\ref{spGpaab}) \\
G_{+\a\b}& (\ref{sGpaa}) \\
G_{+\bar\a\bar\b}& (\ref{sGpabab}) \\
G_{-\a}{}^\a& (\ref{sGmtr})\\
G_{-\bar\a\bar\b}& (\ref{sGmabab}) \\
G_{\a\b\g}& (\ref{sGaaax}) \\
G_{-+\bar\a}& (\ref{sGmpabx}) \\
G_{\a\bar\b\bar\g}& (\ref{sGaabab}) \\
G_{\bar\a\bar\b\bar\g}& (\ref{sGabababx}) \\
G_{-+\a}& (\ref{sGpma}) \\
G_{\a\b\bar\g}& (\ref{sGaaab})
\\ \hline
\end{array}
\end{equation}

The $F$ fluxes are determined by the following equations
\begin{equation}
\begin{array}{|c|c|}\hline
\mathrm{Fluxes } & \mathrm{Equations}
 \\\hline
F_{+\a\b\g\d}& (\ref{selfdual}) \\
F_{+\a\bar\b\g}{}^\g  & (\ref{sFpaacontr}) \\
F_{+\a\b\g\bar\d}& (\ref{symcond}) \\
F_{-\a}{}^\a{}_\b{}^\b& (\ref{sFmtrtr}) \\
F_{-+\a\b\g}& (\ref{Fpmaaa2})
\\ \hline
\end{array}
\end{equation}
The fluxes that do not appear in the above tables are not restricted by the Killing
spinor equations. $F$ is further restricted by the self-duality condition.
The conditions on the geometry have been summarized in section four.

\appendix{The solution of Killing spinor equations for the pure spinor
$\eta=1$ }\la{psolution}

We first begin with the algebraic Killing spinor equations. In particular (\ref{dthree}) and (\ref{dfour}) imply that
\bea
P_+=0~,~~~~~~~~~~~G_{+\a}{}^\a=0
\la{pPpGptr}
\eea
and (\ref{dfive}) implies that
\bea
G_{+\bar\a\bar\b}=0~.
\la{pGpabab}
\eea
The last two equations, (\ref{done}) and (\ref{dtwo}), give
\bea
G_{-+\bar\a}=-G_{\bar\a\b}{}^\b
\la{pGmpabGabtrx}
\eea
and
\bea
P_\a+{1\over 12} \epsilon_{\a}{}^{\bar\b_1\bar\b_2\bar\b_3}
G_{\bar\b_1\bar\b_2\bar\b_3}=0~,
\la{pPa}
\eea
respectively.

The trace of the dual of (\ref{abfive}) gives no  new conditions
because of (\ref{pPpGptr})  and the duality condition
\bea
F_{+\a_1\a_2\a_3\a_4}=0~.
\la{pFpaaaa}
\eea
Substitute this back into (\ref{abfive}) to find
\bea
G_{+\a\bar\b}=0~.
\la{pGpaab}
\eea
Furthermore (\ref{afour}) implies that
\bea
F_{+\a\bar\b\d}{}^\d= -i\Omega_{\a,+\bar\b}~,
\la{pFpaabtr}
\eea
the reality condition of $F$ requires that
\bea
\Omega_{\a,+\bar\b}=-\Omega_{\bar\b,+\a}~,
\eea
and the duality of $F$ leads to
\bea
\Omega_{\a,+}{}^\a=0~.
\eea

The condition (\ref{pthree}) does not give any additional
restrictions on the geometry or the fluxes. Separating (\ref{pone})
into real and imaginary pieces, we find that
\bea
\partial_+\log h+{1\over2} \Omega_{+,-+}=0
\eea
and
\bea
i \Omega_{+,\g}{}^\g+Q_+=0~.
\eea

Next consider the equations (\ref{afive}) and (\ref{abfour}).
Dualize (\ref{afive})  and symmetrize
the free indices to find
\bea
F_{+\bar\g_1\bar\g_2\bar\g_3(\a} \epsilon_{\b)}{}^{\bar\g_1\bar\g_2\bar\g_3}=0~,
\la{psymcondx}
\eea
and use the duality of $F$ to get
\bea
F_{+\a\b\g\bar\d}=0~.
\la{psymcond}
\eea
The remaining equation gives
\bea
G_{+\a\b}=0~.
\la{pGpaa}
\eea
The (\ref{abfour}) implies
\bea
\Omega_{\a,\b+}=0~.
\eea
Equation (\ref{afive}) gives no new conditions.
Next we turn our attention to (\ref{ptwo}).
 Substituting (\ref{psymcond}) and (\ref{pGpaa}), we find that
\bea
\Omega_{+,\a\b}=0~.
\eea

Taking the real and imaginary parts of (\ref{mone}), we find
\bea
\partial_-\log h+{1\over2} \Omega_{-,-+}=0
\eea
and
\bea
-i Q_-+ \Omega_{-,\g}{}^\g+{i\over2} F_{-\g}{}^\g{}_\d{}^\d=0~.
\la{pFmdcontr}
\eea
The latter equation  expresses the $F_{-\g}{}^\g{}_\d{}^\d$ component
of the flux in terms of the geometry
and the scalars.
The condition (\ref{mthree}) gives
\bea
G_{-\g}{}^\g={i\over3} F_{-\bar\b_1\bar\b_2\bar\b_3\bar\b_4}
\epsilon^{\bar\b_1\bar\b_2\bar\b_3\bar\b_4}~,
\la{pGmtr}
\eea
and the condition (\ref{mtwo}) gives
\bea
G_{-\a\b}=\epsilon_{\a\b}{}^{\bar\g_1\bar\g_2} [ \Omega_{-\bar\g_1\bar\g_2}
+ i F_{-\bar\g_1\bar\g_2\d}{}^\d]~.
\la{pGmaa}
\eea

Next let us consider (\ref{atwo}), (\ref{abone}), (\ref{abthree}) and (\ref{mfour}).
First take the trace of (\ref{atwo}) to find
\bea
\Omega_{\b,}{}^\b{}_{\bar\a}- i F_{-+\bar\a\b}{}^\b
-{1\over8} G_{\g_1\g_2\g_3} \epsilon^{\g_1\g_2\g_3}{}_{\bar\a}=0
\la{pFabtrtrx}
\eea
and (\ref{mfour}) gives
\bea
\Omega_{-,+\bar\a}+ i F_{-+\bar\a\b}{}^\b+{1\over24}
 G_{\g_1\g_2\g_3} \epsilon^{\g_1\g_2\g_3}{}_{\bar\a}=0~.
\eea
These give
\bea
\Omega_{\b,}{}^\b{}_{\bar\a}
+ 3\Omega_{-,+\bar\a}+2i F_{-+\bar\a\b}{}^\b=0
\la{pFmpabtrx}
\eea
and
\bea
 G_{\g_1\g_2\g_3} \epsilon^{\g_1\g_2\g_3}{}_{\bar\a}
 =-24[\Omega_{-,+\bar\a}+ i F_{-+\bar\a\b}{}^\b]=12[\Omega_{-,+\bar\a}
 +\Omega_{\b,}{}^\b{}_{\bar\a}]~.
 \la{pGaaax}
\eea
Next eliminating
$G_{\g_1\g_2\g_3}$ from (\ref{abone}), and using (\ref{pFabtrtrx})
 and (\ref{pGaaax}), we get
\bea
D_{\bar\a} \log h+ {1\over2} \Omega_{\bar\a,\b}{}^\b
+ {1\over2} \Omega_{\bar\a,-+}-{1\over4} \Omega_{\b,}{}^\b{}_{\bar \a}
-{7\over4} \Omega_{-,+\bar\a}- {3i\over2} F_{-+\bar\a\b}{}^\b=0~.
\la{pabx}
\eea
In addition, (\ref{abthree}) gives
\bea
G_{\bar\a\g}{}^\g=G_{\bar\a-+}~.
\eea
Comparing this with (\ref{pGmpabGabtrx}), we find that
\bea
G_{\bar\a\g}{}^\g=G_{\bar\a-+}=0~.
\la{pGmpabGabtrx2}
\eea
Finally, dualize (\ref{atwo}) and use (\ref{pGaaax}) to find
\bea
[\Omega_{\a, \bar\g_1\bar\g_2}+2 i
F_{-+\a\bar\g_1\bar\g_2}] \epsilon^{\bar\g_1\bar\g_2}{}_{ \b_1\b_2}
- 2 \Omega_{-,+\bar\g}
\epsilon^{\bar\g}{}_{\a\b_1\b_2}=0~.
\eea
Dualizing the above equation, we get
\bea
\Omega_{\a,\bar\b_1\bar\b_2}+2 i F_{-+\a\bar\b_1\bar\b_2}
-2 \Omega_{-,+[\bar\b_1} g_{\bar\b_2]\a}=0
\la{pFaababtr}
\eea
and taking the trace we find
\bea
\Omega_{\b,}{}^\b{}_{\bar\a}+2i
F_{-+\bar\a\b}{}^\b +3\Omega_{-,+\bar\a}=0~,
\la{pFmpabtry}
\eea
which is identical to (\ref{pFmpabtrx}).

Next we turn our attention to the remaining four equations
(\ref{aone}), (\ref{athree}), (\ref{abtwo}) and (\ref{mfive}).
First (\ref{athree}) implies that
\bea
G_{\a\g}{}^\g- G_{-+\a}- {2i\over3} F_{\a\bar\g_1\bar\g_2\bar\g_3\bar\g_4} \epsilon^{\bar\g_1\bar\g_2\bar\g_3\bar\g_4}=0~.
\la{pGatrGmpax}
\eea
Dualizing and taking the trace of (\ref{abtwo}), we get
\bea
{1\over2} \Omega_{\bar\g_1,\bar\g_2\bar\g_3} \epsilon^{\bar\g_1\bar\g_2\bar\g_3}{}_{\a}
+{1\over8} G_{\a\b}{}^\b+{3\over8} G_{\a-+}=0
\eea
and dualizing (\ref{mfive}), we find
\bea
{i\over3}
F_{-+\bar\g_1\bar\g_2\bar\g_3}\epsilon^{\bar\g_1\bar\g_2\bar\g_3}{}_{\a}
-{1\over8} G_{\a\b}{}^\b-{3\over8} G_{\a-+}=0
\la{pGatrGmpay}
\eea
Adding the above two equations, we get
\bea
3\Omega_{[\bar\g_1,\bar\g_2\bar\g_3]}+ 2i F_{-+\bar\g_1\bar\g_2\bar\g_3}=0~,
\la{pFaaatr}
\eea
which expresses $F_{-+\bar\g_1\bar\g_2\bar\g_3}$ in terms of the  geometry.
In addition comparing (\ref{pGatrGmpax}) and  (\ref{pGatrGmpay}), and using
the duality of $F$, we get
\bea
G_{\a-+}=0
\la{pGpma}
\eea
and
\bea
G_{\a\b}{}^\b={8i\over3}
F_{-+\bar\g_1\bar\g_2\bar\g_3}
\epsilon^{\bar\g_1\bar\g_2\bar\g_3}{}_\a=-4i \Omega_{\g_1, \bar\g_2\bar\g_3}
\epsilon^{\bar\g_1\bar\g_2\bar\g_3}{}_\a~.
\eea
By substituting the above results into (\ref{abtwo}), we also find
\bea
G_{\g_1\g_2\bar\a}=(\Om_{\bar\a,\bar\d_1\bar\d_2}-2i
F_{-+\bar\a\bar\d_1\bar\d_2})\ep^{\bar\d_1\bar\d_2}{}_{\g_1\g_2}=
2[\Omega_{\bar\a,\bar\d_1\bar\d_2}+ \Omega_{\bar\d_1,\bar\d_2\bar\a}]
\ep^{\bar\d_1\bar\d_2}{}_{\g_1\g_2}~.
\la{pGaaab}
\eea

Finally taking the complex conjugate of (\ref{aone}), we get
\bea
{ D^*}_{\bar\a}\log h-{1\over2} \Omega_{\bar\a,\b}{}^\b+{1\over2}
\Omega_{\bar\a,-+}+\frac{1}{4}\Om_{\b,}{}^\b{}_{\bar\a}+\frac{3}{4}\Om_{-,+\bar\a}
+{3i\over2} F_{-+\bar\a\b}{}^\b=0
\eea
Comparing this with (\ref{pabx}) we find
\bea
2\partial_{\bar\a}\log h-\Omega_{\bar\a,+-}-\Omega_{-,+\bar\a}=0
\eea
and
\bea
i Q_{\bar\a}-\Omega_{\bar\a,\b}{}^\b{}+{1\over2}
\Omega_{\b,}{}^\b{}_{\bar\a}& +&{5\over2} \Omega_{-,+\bar\a}
\cr
+3i
F_{-+\bar\a\g}{}^\g&=&
i Q_{\bar\a}-\Omega_{\bar\a,\b}{}^\b{}-
\Omega_{\b,}{}^\b{}_{\bar\a} -2 \Omega_{-,+\bar\a}=0~.
\la{pFpmatr}
\eea

The scalar fluxes $P$ are given in the following equations
\begin{equation}
\begin{array}{|c|c|}\hline
\mathrm{Fluxes } & \mathrm{Equations}
 \\
 \hline
P_+&(\ref{pPpGptr})  \\
P_\a   &  (\ref{pPa})\\
P_{\bar\a}& (\ref{done})
\\ \hline
\end{array}
\end{equation}
The $G$ fluxes are determined by the following equations
\begin{equation}
\begin{array}{|c|c|}\hline
\mathrm{Fluxes } & \mathrm{Equations}
 \\\hline
G_{+\a}{}^\a&(\ref{pPpGptr})  \\
G_{+\a\bar\b}   &  (\ref{pGpaab}) \\
G_{+\a\b}& (\ref{pGpaa}) \\
G_{+\bar\a\bar\b}& (\ref{pGpabab}) \\
G_{-\a}{}^\a& (\ref{pGmtr})\\
G_{-\a\b}& (\ref{pGmaa}) \\
G_{\a\b\g}& (\ref{pGaaax}) \\
G_{-+\bar\a}& (\ref{pGmpabGabtrx2}) \\
G_{-+\a}& (\ref{pGpma}) \\
G_{\a\b\bar\g}& (\ref{pGaaab})
\\ \hline
\end{array}
\end{equation}

The $F$ fluxes are determined by the following equations
\begin{equation}
\begin{array}{|c|c|}\hline
\mathrm{Fluxes } & \mathrm{Equations}
 \\\hline
F_{+\a\b\g\d}& (\ref{selfdual}) \\
F_{+\a\bar\b\g}{}^\g  & (\ref{pFpaabtr}) \\
F_{+\a\b\g\bar\d}& (\ref{psymcond}) \\
F_{-\a}{}^\a{}_\b{}^\b& (\ref{pFmdcontr}) \\
F_{-+\a\b\g} & (\ref{pFaaatr}) \\
F_{-+\a\bar\b\bar\g} & (\ref{pFaababtr})
\\ \hline
\end{array}
\end{equation}
The fluxes that do not appear in the above tables are not restricted
by the Killing spinor equations. $F$ is further restricted by the
self-duality condition. The conditions on the geometry are summarized in
section \ref{pcondition}.

\appendix{The solution of Killing spinor equations for the pure
spinor $\eta=e_{1234}$}\la{p2solution}

We first begin with the algebraic Killing spinor equations. In
particular (\ref{dthree}) and (\ref{dfour}) imply that
\bea
P_+=0~,~~~~~~~~~~~G_{+\a}{}^\a=0
\la{p2PpGptr}
\eea
and (\ref{dfive}) implies that
\bea
G_{+\a\b}=0~.
\la{p2Gpabab}
\eea
The last two equations, (\ref{done}) and (\ref{dtwo}), give
\bea
G_{-+\a}=G_{\a\b}{}^\b
\la{p2GmpabGabtrx}
\eea
and
\bea
P_{\bar\a}+{1\over 12} \epsilon_{\bar\a}{}^{\b_1\b_2\b_3}
G_{\b_1\b_2\b_3}=0~,
\eea
respectively.

The trace of the dual of (\ref{afour}) gives the duality condition
\bea
F_{+\a_1\a_2\a_3\a_4}=0~.
\la{p2Fpaaaa}
\eea
Substitute this back into (\ref{afour}) to find
\bea
G_{+\a\bar\b}=0~.
\la{p2Gpaab}
\eea
Furthermore (\ref{abfive}) implies that
\bea
F_{+\a\bar\b\d}{}^\d= i\Omega_{\a,+\bar\b}~,
\la{p2Fpaabtr}
\eea
the reality condition of $F$ requires that
\bea
\Omega_{\a,+\bar\b}=-\Omega_{\bar\b,+\a}~,
\eea
and the duality condition of $F$ gives
\bea
\Omega_{\a,+}{}^\a=0~.
\eea
The condition (\ref{pone}) does not give any additional restrictions
on the geometry or the fluxes. Separating (\ref{pthree}) into real
and imaginary pieces, we find that
\bea
\partial_+\log h+{1\over2} \Omega_{+,-+}=0
\eea
and
\bea
-i \Omega_{+,\g}{}^\g+Q_+=0~.
\eea

Next consider the equations (\ref{afive}),  (\ref{abfour}) and use the duality
of $F$. They
yield
\bea
F_{+\a\b\g\bar\d}=0~,
\la{p2Fpaaaab}
\eea
\bea
 \Omega_{\bar\a,\bar\b+}=0~,
\eea
and
\bea
G_{+\bar\a\bar\b}=0~.
\la{p2Gpaa}
\eea

Next we turn our attention to (\ref{ptwo}). Using the above results,
we find that
\bea
\Omega_{+,\a\b}=0~.
\eea

Taking the real and imaginary parts of (\ref{mthree}), we find
\bea
\partial_-\log h+{1\over2} \Omega_{-,-+}=0
\eea
and
\bea
-i Q_- - \Omega_{-,\g}{}^\g+{i\over2} F_{-\g}{}^\g{}_\d{}^\d=0~.
\eea
The latter equation can expresses the $F_{-\g}{}^\g{}_\d{}^\d$
component of the flux in terms of the geometry and the scalars. The
condition (\ref{mone}) gives
\bea
G_{-\g}{}^\g=-{i\over3} F_{-\b_1\b_2\b_3\b_4}
\epsilon^{\b_1\b_2\b_3\b_4}~,
\eea
and the condition (\ref{mtwo}) gives
\bea
G_{-\bar\a\bar\b}=\epsilon_{\bar\a\bar\b}{}^{\g_1\g_2} [
\Omega_{-,\g_1\g_2}- i F_{-\g_1\g_2\d}{}^\d]~.
\eea

Next let us consider (\ref{aone}), (\ref{athree}), (\ref{abtwo}) and
(\ref{mfive}). An analysis similar to the one in the previous
appendix yields
\bea
\Omega_{\b,}{}^\b{}_{\bar\a} +
3\Omega_{-,+\bar\a}-2i F_{-+\bar\a\b}{}^\b=0
\la{p2Fmpabtrx}
\eea
and
\bea
 G_{\bar\g_1\bar\g_2\bar\g_3} \epsilon^{\bar\g_1\bar\g_2\bar\g_3}{}_{\a}
 =-24[\Omega_{-,+\a}- i F_{-+\a\b}{}^\b]=12
 [\Omega_{-,+\a}+\Omega_{\bar\b,}{}^{\bar\b}{}_\a]~.
 \la{p2Gaaax}
\eea
The former condition has been viewed as an equation for
$F_{-+\bar\a\b}{}^\b$. By substituting these results back
into the equations we find
\bea
 D^*_{\bar\a} \log h+ {1\over2} \Omega_{\bar\a,\b}{}^\b+ {1\over2}
\Omega_{\bar\a,-+}-{1\over4} \Omega_{\b,}{}^\b{}_{\bar \a} -{7\over4}
\Omega_{-,+\bar\a}+ {3i\over2} F_{-+\bar\a\b}{}^\b=0~.
\la{yyxx}
\eea
and
\bea
G_{\a\g}{}^\g=G_{\a-+}=0~.
\eea
Finally, dualizing (\ref{abtwo}) yields
\bea
\Omega_{\bar\a,\b_1\b_2}+2 i
F_{-+\bar\a\b_1\b_2} -2 \Omega_{-,+[\b_1} g_{\b_2]\bar\a}=0~.
\eea

Next we turn our attention to the remaining four equations
(\ref{atwo}), (\ref{abone}), (\ref{abthree}) and (\ref{mfour}). They
yield
\bea
3\Omega_{[\bar\g_1,\bar\g_2\bar\g_3]}- 2i
F_{-+\bar\g_1\bar\g_2\bar\g_3}=0~,
\eea
which expresses $F_{\bar\g_1\bar\g_2\bar\g_3\d}{}^\d$ in terms of
the other fluxes and geometry. We also find that
\bea
G_{\bar\a-+}=0
\eea
and
\bea
G_{\bar\a\b}{}^\b=-{8i\over3} F_{-+\g_1\g_2\g_3}
\epsilon^{\g_1\g_2\g_3}{}_{\bar\a}=4\Omega_{\g_1,\g_2\g_3}
\epsilon^{\g_1\g_2\g_3}{}_{\bar\a}~.
\eea
By substituting the above results into (\ref{atwo}) we get
\bea
G_{\a\bar\g_1\bar\g_2}&=&(\Om_{\a,\d_1\d_2}-2i
F_{-+\a\d_1\d_2})\ep^{\d_1\d_2}{}_{\bar\g_1\bar\g_2}
\cr
&=&2[\Omega_{\a,\d_1\d_2}+\Omega_{\d_1,\d_2\a}]\ep^{\d_1\d_2}{}_{\bar\g_1\bar\g_2}~.
\eea
Finally, comparing (\ref{abthree}) with (\ref{yyxx})
yields
\bea
2\partial_{\bar\a}\log h-\Omega_{\bar\a,+-}-\Omega_{-,+\bar\a}=0
\eea
and
\bea
-i Q_{\bar\a}-\Omega_{\bar\a,\b}{}^\b{}+{1\over2}
\Omega_{\b,}{}^\b{}_{\bar\a} &+&{5\over2} \Omega_{-,+\bar\a}
\cr
-3i
F_{-+\bar\a\g}{}^\g&=&
-i Q_{\bar\a}-\Omega_{\bar\a,\b}{}^\b{}-
\Omega_{\b,}{}^\b{}_{\bar\a} -2 \Omega_{-,+\bar\a}=0~.
\eea
This concludes the analysis of the equations. The conditions on the geometry
 are
summarized in section \ref{pcondition}.

\end{document}